\documentclass[twocolumn]{aastex62}

\usepackage{mathtools}
\usepackage{chngcntr}
\usepackage{footnote}
\usepackage[caption=false]{subfig}
\usepackage{float}
\usepackage{tabulary,booktabs}

\usepackage{multirow}
\usepackage{graphicx}
\usepackage{amsmath}
\usepackage{yhmath}
\usepackage{mathtools}
\usepackage{csquotes}

\def\farmer{\texttt{The Farmer}}
\def\Euclid{\textit{Euclid}}
\def\lephare{\texttt{LePhare}}

\newcommand{\new}[1]{{#1}}

\shorttitle{A machine learning approach to predict missing flux densities}
\shortauthors{Chartab et al.}

\begin{document}

\title{\textbf{A Machine Learning Approach to Predict Missing Flux Densities in Multi-band Galaxy Surveys}}

\correspondingauthor{Nima Chartab}
\email{nchartab@carnegiescience.edu}

\author[0000-0003-3691-937X]{Nima Chartab}
\affiliation{The Observatories of the Carnegie Institution for Science, 813 Santa Barbara St., Pasadena, CA 91101, USA}
\affiliation{Department of Physics and Astronomy, University of California, Irvine, CA 92697, USA}
\affil{Department of Physics and Astronomy, University of California, Riverside, 900 University Ave, Riverside, CA 92521, USA}

\author{Bahram Mobasher}
\affil{Department of Physics and Astronomy, University of California, Riverside, 900 University Ave, Riverside, CA 92521, USA}

\author{Asantha R. Cooray}
\affiliation{Department of Physics and Astronomy, University of California, Irvine, CA 92697, USA}

\author[0000-0003-2226-5395]{Shoubaneh Hemmati}
\affiliation{Infrared Processing and Analysis Center, California Institute of Technology, Pasadena, CA 91125, USA}

\author[0000-0002-0364-1159]{Zahra Sattari}
\affil{Department of Physics and Astronomy, University of California, Riverside, 900 University Ave, Riverside, CA 92521, USA}
\affiliation{The Observatories of the Carnegie Institution for Science, 813 Santa Barbara St., Pasadena, CA 91101, USA}

\author{Henry C. Ferguson}
\affiliation{Space Telescope Science Institute, 3700 San Martin Drive, Baltimore, MD 21218, USA}

\author{David B. Sanders}
\affiliation{Institute for Astronomy (IfA), University of Hawaii, 2680 Woodlawn Drive, Honolulu, HI 96822, USA} 

\author[0000-0003-1614-196X]{John R. Weaver}
\affiliation{Cosmic Dawn Center (DAWN)}
\affiliation{Niels Bohr Institute, University of Copenhagen, Jagtvej 128, DK-2200 Copenhagen, Denmark}

\author{Daniel K. Stern}
\affiliation{Jet Propulsion Laboratory, California Institute of Technology, 4800 Oak Grove Drive, Pasadena, CA 91109, USA}

\author{Henry J. McCracken}
\affiliation{Institut d'Astrophysique de Paris, UMR 7095, CNRS, and Sorbonne Universit\'e, 98 bis boulevard Arago, 75014 Paris, France}

\author{Daniel C. Masters}
\affiliation{Infrared Processing and Analysis Center, California Institute of Technology, Pasadena, CA 91125, USA}

\author{Sune Toft}
\affiliation{Cosmic Dawn Center (DAWN)}
\affiliation{Niels Bohr Institute, University of Copenhagen, Jagtvej 128, DK-2200 Copenhagen, Denmark}

\author{Peter L. Capak}
\affiliation{Infrared Processing and Analysis Center, California Institute of Technology, Pasadena, CA 91125, USA}

\author{Iary Davidzon}
\affiliation{Cosmic Dawn Center (DAWN)}
\affiliation{Niels Bohr Institute, University of Copenhagen, Jagtvej 128, DK-2200 Copenhagen, Denmark}

\author{Mark E. Dickinson}
\affiliation{National Optical Astronomy Observatories, 950 N Cherry
Avenue, Tucson, AZ 85719, USA}

\author{Jason Rhodes}
\affiliation{Jet Propulsion Laboratory, California Institute of Technology, 4800 Oak Grove Drive, Pasadena, CA 91109, USA}

\author{Andrea Moneti}
\affiliation{Institut d'Astrophysique de Paris, UMR 7095, CNRS, and Sorbonne Universit\'e, 98 bis boulevard Arago, 75014 Paris, France}

\author{Olivier Ilbert}
\affiliation{Aix Marseille Univ, CNRS, LAM, Laboratoire d'Astrophysique de Marseille, Marseille, France}

\author{Lukas Zalesky}
\affiliation{Institute for Astronomy (IfA), University of Hawaii, 2680 Woodlawn Drive, Honolulu, HI 96822, USA}

\author{Conor J.R. McPartland}
\affiliation{Institute for Astronomy (IfA), University of Hawaii, 2680 Woodlawn Drive, Honolulu, HI 96822, USA}

\author{Istv\'an Szapudi}
\affiliation{Institute for Astronomy (IfA), University of Hawaii, 2680 Woodlawn Drive, Honolulu, HI 96822, USA} 

\author[0000-0002-6610-2048]{Anton M. Koekemoer}
\affiliation{Space Telescope Science Institute, 3700 San Martin Dr.,
Baltimore, MD 21218, USA}

\author{Harry  I. Teplitz}
\affiliation{Infrared Processing and Analysis Center, California Institute of Technology, Pasadena, CA 91125, USA} 

\author{Mauro Giavalisco}
\affiliation{Department of Astronomy, University of Massachusetts, 710 North Plesant Street, Amherst, MA 01003, USA}

\begin{abstract}
\label{abstract}
We present a new method based on information theory to find the optimal number of bands required to measure the physical properties of galaxies with a desired accuracy. As a proof of concept, using the recently updated COSMOS catalog (COSMOS2020), we identify the most relevant wavebands for measuring the physical properties of galaxies in a Hawaii Two-0 (H20)- and UVISTA-like survey for a sample of $i<25$ AB mag galaxies. We find that with available $i$-band fluxes, $r$, $u$, IRAC/$ch2$ and $z$ bands provide most of the information regarding the redshift with importance decreasing from $r$-band to $z$-band. We also find that for the same sample, IRAC/$ch2$, $Y$, $r$ and $u$ bands are the most relevant bands in stellar mass measurements with decreasing order of importance. Investigating the inter-correlation between the bands, we train a model to predict UVISTA observations in near-IR from H20-like observations. We find that magnitudes in $YJH$ bands can be simulated/predicted with an accuracy of $1\sigma$ mag scatter $\lesssim 0.2$ for galaxies brighter than 24 AB mag in near-IR bands. One should note that these conclusions depend on the selection criteria of the sample. For any new sample of galaxies with a different selection, these results should be remeasured. Our results suggest that in the presence of a limited number of bands, a machine learning model trained over the population of observed galaxies with extensive spectral coverage outperforms template-fitting. Such a machine learning model maximally comprises the information acquired over available extensive surveys and breaks degeneracies in the parameter space of template-fitting inevitable in the presence of a few bands. 
\end{abstract}


\section{Introduction}

\label{sec:Introduction}

Future ground-based and space-borne observatories, equipped with large aperture telescopes and sensitive large format detectors will provide broad-band imaging data for more than a billion galaxies. These data are pivotal to better understanding of dark sectors of the Universe (i.e., dark matter and dark energy) as well as the evolution of galaxies and large-scale structures over cosmic time. The challenge, however, is to obtain wide waveband coverage to constrain the spectral energy distributions (SEDs) of millions of galaxies and estimate their redshifts and physical parameters such as stellar masses and star formation rates. 

Template fitting is widely used to infer photometric redshifts of galaxies and their physical properties \cite[e.g.,][]{Arnouts99,Bolzonella2000,Ilbert06}. However, theoretical synthetic templates may not be representative of the real parameter space of galaxies. For example, templates can include SEDs which do not have an observational analog. This will cause degeneracy in parameter measurement, especially when we reconstruct SEDs with few bands. Many of these degeneracies are mitigated by obtaining data with wide spectral coverage (e.g., with a larger number of wavebands). An example of such a data set is the Cosmic Evolution Survey \cite[COSMOS;][]{Scoville07} that has been observed in more than 40 bands from X-ray to radio wavelengths. The wealth of information in this field provides very well-constrained SEDs for galaxies. However, not all surveys have as many photometric bands as the COSMOS field. For instance, \Euclid{} \citep{Laureijs11} will rely on near-infrared $Y$, $J$,
and $H$ bands ($960–2000\ \rm nm$), complemented by optical ground-based observations in $u$, $g$, $r$, $i$ and $z$ to measure photometric redshifts \citep{Euclid_photz20}. It is therefore instructive to use the extensive dataset in the COSMOS field to identify essential bands which carry most of the information regarding the physical properties of galaxies. 

The aim of this study is to transfer the information gained in the COSMOS field to fields such as the \Euclid{} deep fields where such extensive photometry does not exist. Using the concepts of information theory, we can find if there is any information shared between the bands and use these measurements to identify the most important bands (those that reveal most of the information about the physical properties of galaxies). Based on the machine learning techniques, we can then predict fluxes in the wavebands that are not observed in a survey but share information with other available (observed) bands. This allows us to carefully design future surveys and only observe in selected wavebands that include most of the information to significantly save in the observing time.

Machine learning has become popular in recent years to build models based on spectroscopic redshifts \citep[e.g.,][]{Carrasco14,Masters17} and train models based on synthetic templates \citep[e.g.,][]{Hemmati19} or mock catalogs generated from galaxy simulations \citep[e.g.,][]{Davidzon19,Simet21}. These methods are particularly useful as machine learning algorithms can learn more complicated relations given a large and comprehensive training data set \citep{Mucesh21}. Moreover, these models speed up parameter measurement, which is  an important characteristic with the flood of data imminent from upcoming surveys \citep{Hemmati19}.

In this paper, we develop a new technique based on information theory to quantify the importance of each waveband and identify essential bands to measure the physical properties of galaxies. We also develop a machine learning model to predict fluxes in missing bands and thereby improve the wavelength resolution of existing photometric data. To demonstrate the application of these techniques, we apply our methods  to a sample of galaxies drawn from the latest version of the COSMOS survey \citep[COSMOS2020;][]{Weaver21}, analogous to that planned by the \Euclid{} deep fields. A new ground-based survey, Hawaii Two-0 (H20; McPartland et al. in prep), has been designed to provide complementary photometric data for the \Euclid{} mission. H20 will provide $u-$band observations from MegaCam instrument on the Canada-France-Hawaii telescope (CFHT) and $g-,r-,i-, z-$band imaging from Hyper Suprime-Cam (HSC) instrument on the Subaru telescope over 20 square degrees of the \Euclid{} deep fields. Spitzer/IRAC observations from the Spitzer Legacy Survey (SLS) are also available in the same fields \citep{Moneti21}. In this paper, we identify the importance of wavebands for an H20+UVISTA-like survey with similar wavelength coverage expected in \Euclid{} deep fields, incorporating the near-IR $YJH$ bands from UltraVista \citep{McCracken12} in addition to the H20 and SLS wavebands. We then predict fluxes in near-IR wavebands using the existing ground-based and mid-IR Spitzer/IRAC observations (H20-like) of the deep fields.

In Section \ref{sec:Data}, we briefly introduce the COSMOS2020 catalog, and use that to build a sample of H20+UVISTA-like galaxies. Section \ref{information} describes the concepts of information gain and quantifies the importance of each waveband based on that. In Section \ref{sec:Visualize}, we use dimensionality reduction techniques to visualize photometric data in 2-dimensional space to explore the feasibility of predicting fluxes in near-IR fluxes based on $ugriz$ and Spitzer/IRAC data. This is followed by Section \ref{sec:band_prediction} where we train a machine learning algorithm, Random Forest model, to predict fluxes in UVISTA/$YJH$ wavebands using data in wavebands similar to the existing H20. In Section \ref{sec:Photz-M}, we investigate the accuracy of the photometric redshifts and stellar masses given the limited number of bands available in H20-like and H20+UVISTA-like data. We discuss and summarize our results in Section \ref{sec:Discussion_Summary}.              

Throughout this work, we assume flat $\Lambda$CDM cosmology with $H_0=70 \rm \ kms^{-1} Mpc^{-1}$, $\Omega_{m_{0}}=0.3$ and $\Omega_{\Lambda_{0}}=0.7$. All magnitudes are expressed in the AB system, and the physical parameters are measured assuming a \cite{Chabrier03} IMF.

\begin{figure}
    \centering
    \includegraphics[width=1\linewidth,clip=True, trim=1.4cm 0cm 2cm 0cm]{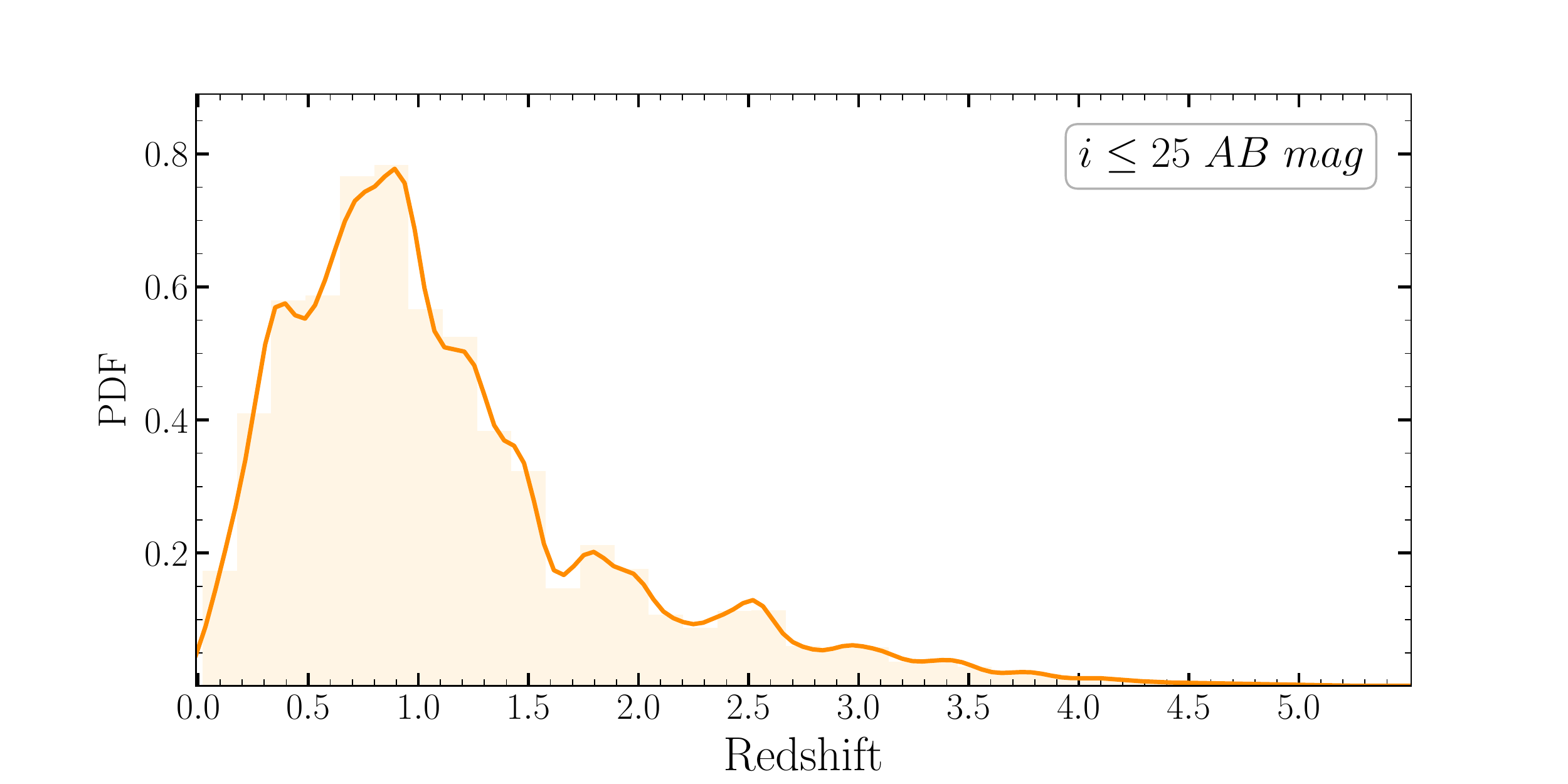}
    \caption{Redshift distribution for the subset of COSMOS2020 galaxies brighter than $i$= 25 AB magnitude (3$\sigma$). The entropy of the redshift calculated based on the distribution shown in this figure is less than the entropy of a uniformly distributed redshift. In other words, we get less surprised when we observe the redshift of a galaxy given this distribution (prior information). }
    \label{fig:z_PDF}
\end{figure}

\section{Data}
\label{sec:Data}

Here we use the updated version of the COSMOS catalog, COSMOS2020, to build a sample of galaxies analogous to those that will be observed in the \Euclid{} deep fields. Compared to COSMOS2015 catalog \citep{Laigle16}, COSMOS2020 provides much deeper near-IR and mid-IR (Spitzer) photometric data as well as two independent methods for photometric extraction - the conventional and a profile-fitting (\farmer{}; J. Weaver et al., in prep.) methods. We use \farmer{} photometry that contains consistent photometric data in 39 bands from FUV to mid-IR including broad, medium and narrow filters. All the data are reduced to the same scale with appropriate PSFs. Photometric redshifts are calculated using LePhare \citep{Arnouts99,Ilbert06} with a similar configuration described in \cite{Ilbert13}. Given the large number of bands with deep observations, photometric redshift solutions are accurate, reaching a normalized median absolute deviation \citep[$\sigma_{\rm NMAD}$;][]{Hoaglin83} of $0.02$ for galaxies as faint as $i\sim25$ AB mag \citep{Weaver21}. The redshifts of galaxies are then fixed on their estimated photometric redshifts and the stellar masses were estimated. In this paper, we consider COSMOS2020 photometric redshifts and stellar masses as a \enquote{ground truth} since spectroscopic redshifts are only available for a limited number of galaxies and using a mixture of photometric and spectroscopic redshifts can bias our sample towards specific populations of galaxies. 

We use two sets of wavebands: 1) H20-like bands: ${\rm \mathbf{A}}\coloneqq\{u,g,r,i,z,ch1,ch2\}$, 2) H20+UVISTA-like bands: ${\rm \mathbf{B}}\coloneqq\{u,g,r,i,z,Y,J,H,ch1,ch2\}$. $u-$band observations are conducted by MegaCam instrument at CFHT, and other optical bands ($g,r,i$ and $z$) are available from Subaru's Hyper Suprime-Cam (HSC) imaging. Spitzer/IRAC channel 1,2 ($ch1,ch2$) data are compiled from all the IRAC observations of the COSMOS field \citep{Moneti21}. Near-IR photometry in $Y$, $J$ and $H$ bands are obtained from the UltraVista survey \citep{McCracken12}. We select a subset of the COSMOS2020 galaxies that are observed, but not necessarily detected, in all the aforementioned bands and have $i-$band AB magnitude $\leq 25$ with $3\sigma$ detection. These selection criteria result in 165,807 galaxies out to $z\sim 5.5$. \new{Photometric measurements in COSMSOS2020 catalog are not corrected for Galactic extinction. We corrected them using \cite{Schlafly11} dust map. Moreover,} some sources have negative fluxes in the desired bands, which is due to the variation of background flux across the image. We set these fluxes to zero.       

\section{Information Gain}
\label{information}
Let's suppose that we do not have any prior information about the redshift distribution of galaxies selected from the criteria mentioned in Section \ref{sec:Data}. We, therefore, assume a uniform distribution for the redshift. As an example, if we define four bins of redshifts (\{$z_1$=(0,1], $z_2$=(1,2], $z_3$=(2,3], $z_4$=(3,4]\}) and want to identify which bin does a galaxy belong to, we can encode it in two bits, as below, \\       



\begin{center}
    \includegraphics[]{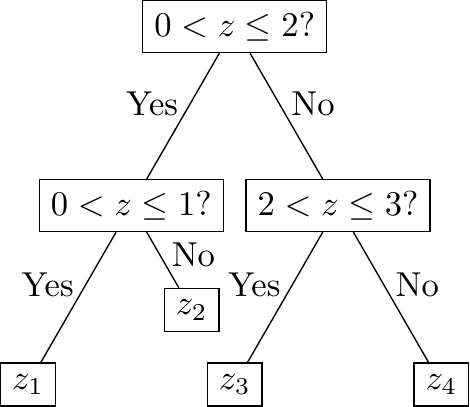}

\end{center}

\begin{figure*}
    \centering
    \includegraphics[width=1\linewidth,clip=True]{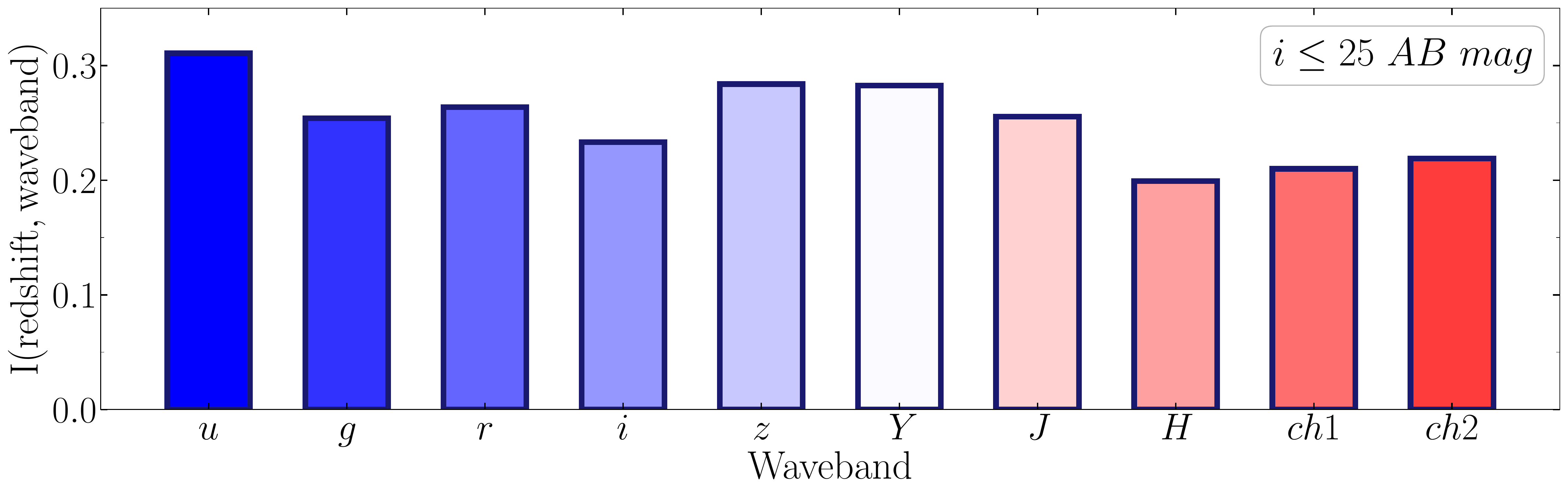}
    \caption{Mutual information of redshift and wavebands in bits per galaxy.  Larger mutual information means that the entropy of the redshift will decrease more if we include the band in photometric redshift measurements, so the band is more important. Here, $u$ is the most important followed by $z$-band.
}
    \label{fig:mi}
\end{figure*}

Here, we need to ask two YES/NO questions to identify the bin a galaxy belongs to. However, based on the available observations of COSMOS2020, we know the redshift distribution of galaxies with $i\leq 25$ AB mag as background information. We, therefore, update the decision tree above, considering our prior information about the redshift distribution, to reduce the average number of questions we need to ask to identify the redshift bin of a galaxy. Based on the redshift distribution shown in Figure \ref{fig:z_PDF}, the probability of a galaxy being in each redshift bin is: $P(z_1)=0.56, P(z_2)=0.32, P(z_3)=0.09, P(z_4)=0.03$. Thus, one possible decision tree to identify the redshift bin of a galaxy can be built as follows,



\begin{center}
    \includegraphics[]{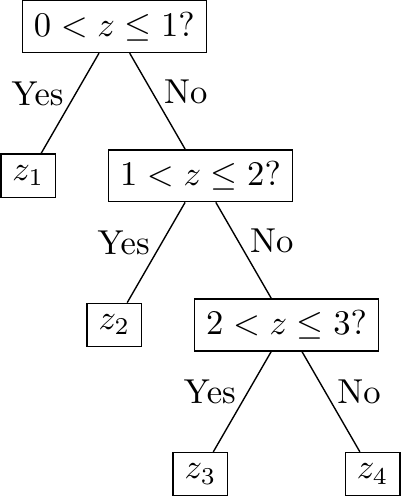}

\end{center}

\noindent On average, $0.56\times 1+0.32\times 2+(0.09+0.03)\times 3=1.56$ questions (bits) are required to identify the redshift bin of a galaxy. We find that the number of bits (questions) reduced from 2 to 1.56 when we added information regarding the redshift distribution of galaxies. This decrease shows that we will get less surprised when we observe the redshift of a galaxy, given that we know what the redshift distribution looks like. 

Given the above example, the optimal number of bits required to store a variable called Shannon’s entropy ($H$), is defined as \citep{Shannon48},

\begin{equation}
    H(X)=-\sum_i P(x_i)\log_2 P(x_i),
    \label{entropy}
\end{equation}

\begin{figure*}[ht]
    \centering
    \includegraphics[width=0.95\textwidth]{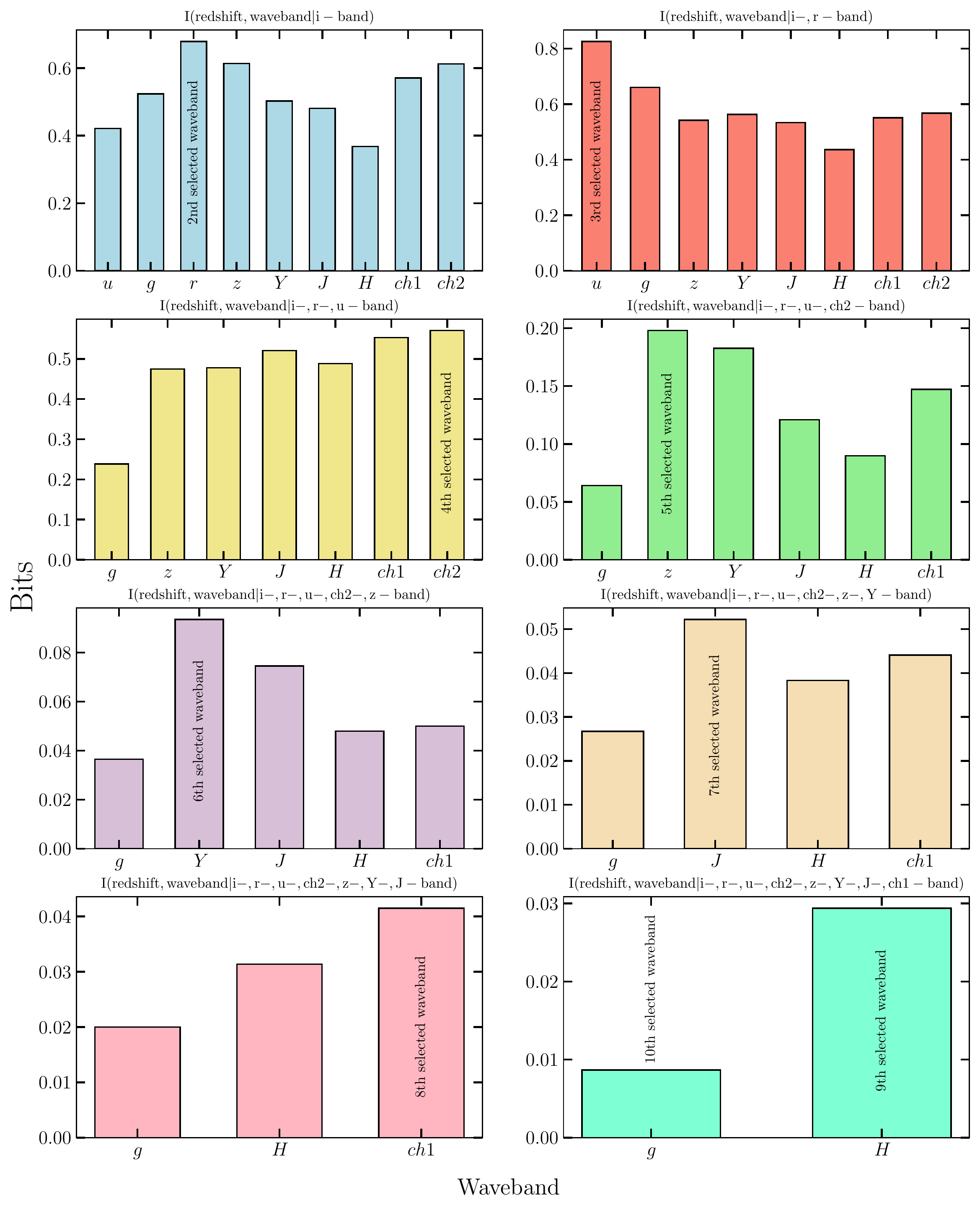}
    \caption{Conditional mutual information of redshift and wavebands in bits per galaxy. The most relevant bands can be selected based on their conditional mutual information. The sample is selected based on the magnitude of the $i-$band, which implies that the first selected waveband is the $i-$band. The top left panel shows the mutual information of redshift and wavebands given that $i-$band data are available. Therefore we select $r-$band as the second most relevant band since it provides the most information. In the top right, we assume that $i-$ and $r-$band data are available and find that $u-$band would be the third choice. We follow a similar procedure to find relevant bands in order of their importance. We note that these results depend on the selection criteria. For any new sample of galaxies with a different selection, these results should be remeasured.    }
    \label{Fig:cmi}
\end{figure*}
\noindent where $x_i$ is the possible outcome of a variable ($X$) which occurs with probability $P(x_i)$. In this formulation, $\log_2 P(x_i)$ represents the number of bits required to identify the outcome. Using equation \ref{entropy}, Shannon’s entropy of redshift based on the probabilities in four bins is 1.45 bits. This means that we can still make our tree more optimal to encode the redshift values in 1.45 bits instead of 1.56. One possible way would be building the tree to identify the redshift of two galaxies simultaneously, which makes the average number of questions per galaxy even less than 1.56. However, we do not aim to find the optimal compression algorithm to encode the redshift information. We just use Shannon’s entropy to find the maximal compression rate. 

\begin{figure*}
    \centering
    \includegraphics[width=1\linewidth,clip=True]{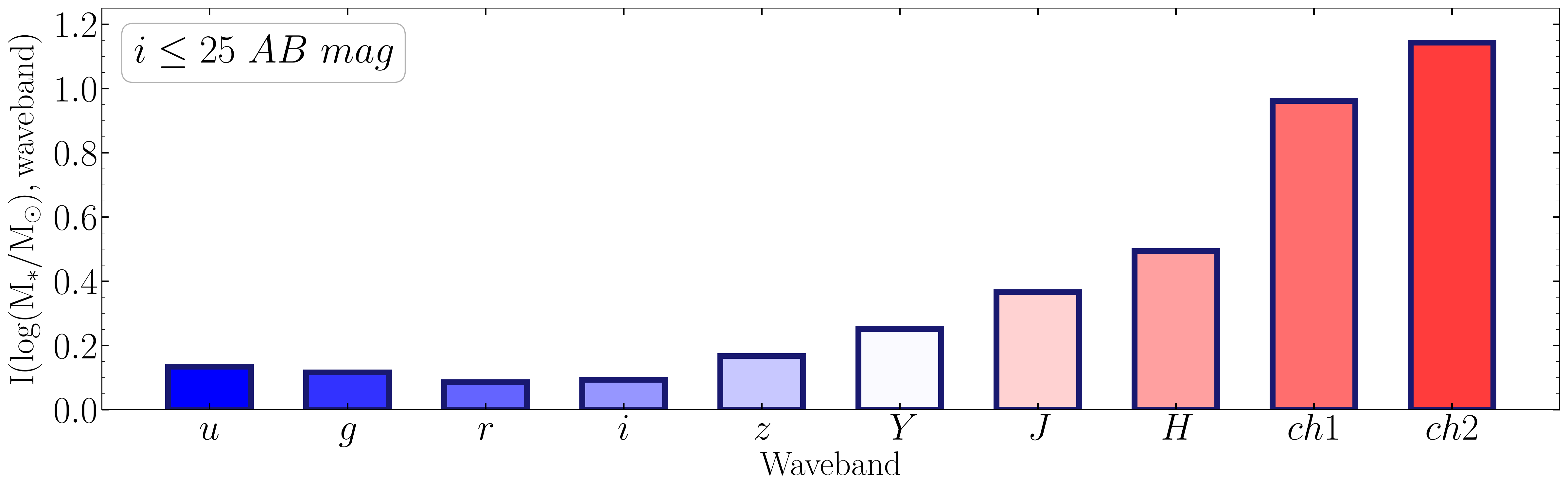}
    \caption{Similar to Figure \ref{fig:mi} but for the stellar mass. Mutual information of stellar mass and wavebands in bits per galaxy is shown. With more mutual information, the entropy of stellar mass will decrease more if we include the band in the photometric stellar mass measurements, so the band is more important.
}
    \label{fig:mi_mass}
\end{figure*}

In the presence of other information, such as observed fluxes in different bands, the entropy of the redshift decreases even more. The amount of uncertainty (entropy) remaining in $X$ after we have seen $Y$ is called conditional entropy and defined as,
\begin{equation}
    H(X|Y)=-\sum_{x\in X,y \in Y} P(x,y)\log_2 \frac{P(x,y)}{P(y)},
\end{equation}where $P(x,y)$ is the joint probability distribution at $(x,y)$. Moreover, the mutual information between X and Y (i.e., the amount of uncertainty in X that is removed by knowing Y) is defined as,

\begin{equation}  \label{eq:mi}
\begin{split}
I(X,Y)&=H(X)-H(X|Y) \\
 & = H(X) + H(Y) - H(X,Y),
\end{split}
\end{equation}where $H(X,Y)$ is the joint entropy of a pair of variables $(X,Y)$. In other words, I$(X,Y)$ is a measure of the amount of information (in bits) one can acquire about $X$ by observing $Y$. This parameter can be used to identify the waveband that will be most useful for measuring galaxy properties (e.g., redshifts). For instance, the waveband with the highest I$(redshift,waveband)$ carries the most information and decreases the entropy of the redshift the most.     

\begin{figure}
    \centering
    \includegraphics[width=0.5\textwidth]{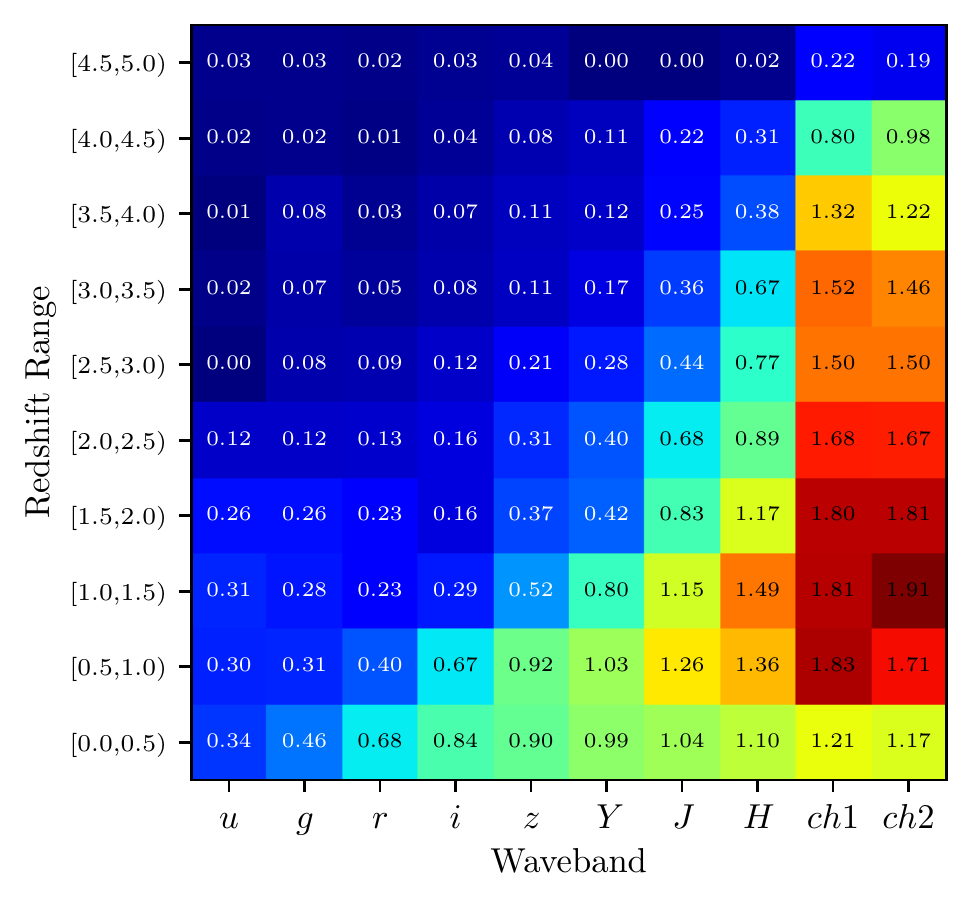}
    \caption{Mutual information of stellar mass and wavebands in bits per galaxy in the bins of redshift. The map is colored based on the value of mutual information, with red representing the most important band and blue representing the least important band. The role of low wavelength bands decreases as we approach higher redshift, as we would expect.}
    \label{fig:mi_bands_z}
\end{figure}
The mutual information as in equation \ref{eq:mi} is defined for discrete variables. In the case of continuous variables (e.g., redshift, flux, stellar mass), we need to properly discretize the data. \cite{Kraskov04} (hereafter KSG) introduced a k-nearest neighbor estimator to compute the mutual information of continuous variables. This method detects the underlying probability distribution of data by measuring distances to the $k^{th}$ nearest neighbors of points in the data set. There is nonzero mutual information when some points are clustered in the X-Y space, which allows us to predict $y\in Y$ given an $x\in X$ coordinate. We refer readers to the original KSG paper for details of the method. Figure \ref{fig:mi} shows the mutual information between redshift and each waveband based on the KSG algorithm with $k=100$ nearest neighbors. It suggests that given the sample of $i<25$ AB mag galaxies, the $u-$band provides the largest information regarding the redshift compared to the rest of the H20+UVISTA-like bands. However, our sample is selected based on $i-$band magnitudes, so we assumed that $i-$band data are already available. Suppose that for our sample $u-$band fluxes are highly correlated with $i-$band data. In this case, $u-$band carries no information in the presence of $i-$band data. To take into account such an effect, we need to compute conditional mutual information, defined as,

\begin{equation}
    I(X,Y|Z)=H(X|Z)- H(X|Y,Z),
    \label{eq:cmi}
\end{equation} where I$(X,Y|Z)$ is the mutual information of $X$ and $Y$ given that $Z$ is observed. Following the KSG algorithm, we find the conditional mutual entropy to sort wavebands based on their importance. We compute I($redshift,waveband|i-band$) and choose the waveband with the highest conditional mutual information as the most important band. The conditional mutual information estimations reveal that the $r-$band is the most important waveband given that $i-$band data are available. We continue computing conditional mutual information, I($redshift,waveband|swaveband$), where $swaveband$ is the previously selected waveband. 

\begin{figure*}[ht]
    \centering
    \includegraphics[width=0.95\textwidth]{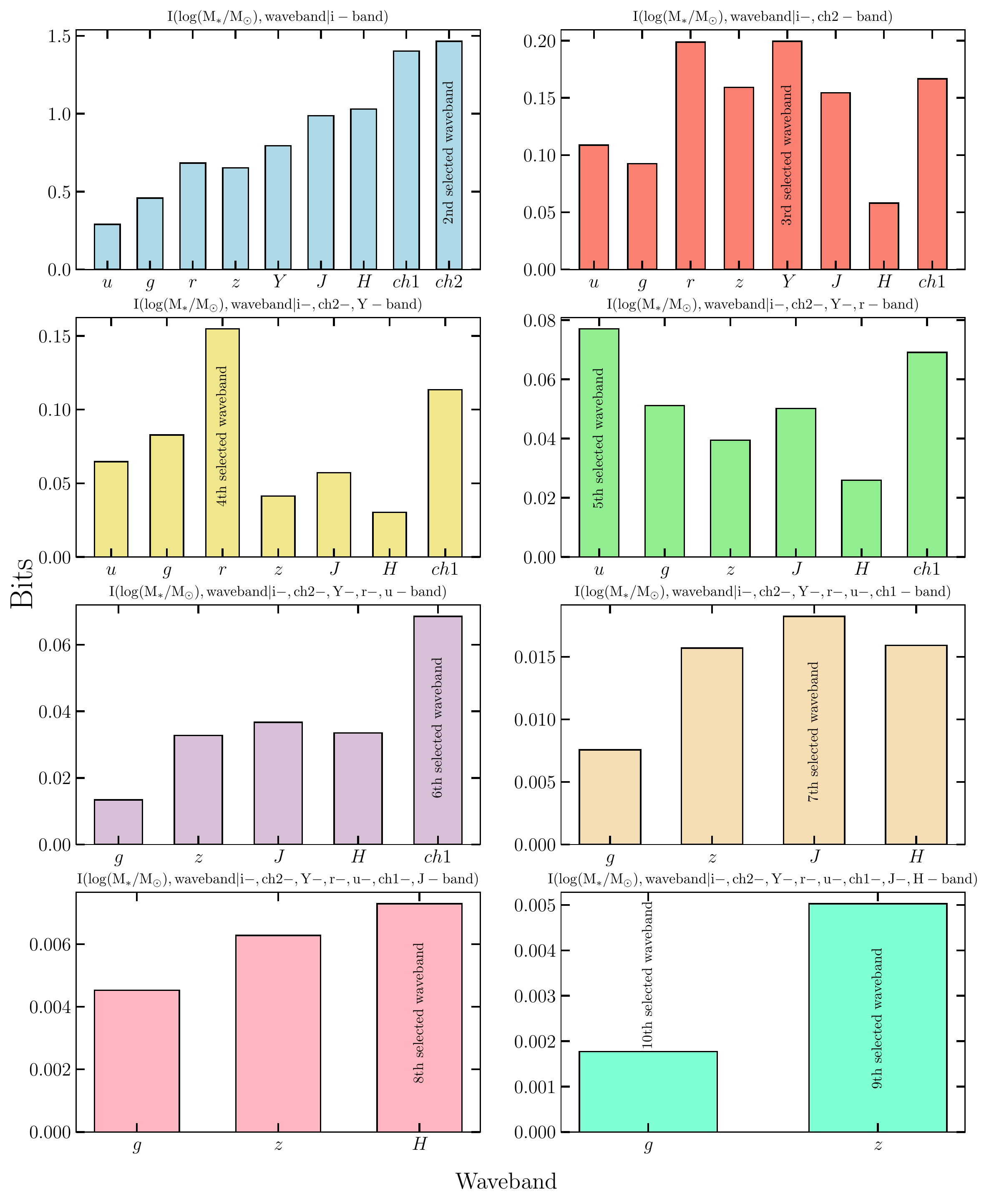}
    \caption{Similar to Figure \ref{Fig:cmi} but for the stellar mass. Each panel shows the Conditional mutual information of stellar mass and wavebands given that all the previously selected bands are available. We find that for the $i-$band selected sample, $ch2$,$Y$,$r$ and $u-$band are the four most relevant bands with decreasing order of importance. The top left panel shows that IRAC data are essential for stellar mass measurements.}
    \label{Fig:cmi_mass}
\end{figure*}

Figure \ref{Fig:cmi} shows the non-zero conditional mutual information as we select relevant wavebands. We find that for $i<25$ AB mag galaxies, $r, u, ch2$ and $z$ bands are the bands that provide most of the information about the redshift with decreasing importance from $r-$band to $z-$band. We repeat these analyses for stellar mass measurements. As shown in Figure \ref{fig:mi_mass}, we measure the mutual information between stellar mass and each waveband for the whole sample, and in Figure \ref{fig:mi_bands_z}, we measure the same quantity, I($\log(M_*/M_\odot),waveband|i-band$), in the bins of redshifts. As we expect, the role of short wavelength bands decreases as we approach higher redshifts. We further compute the important wavebands given the availability of $i-$band data in Figure \ref{Fig:cmi_mass}. We find that $ch2$, $Y$, $r$ and $u$ bands are the most relevant bands in the stellar mass measurements with decreasing order of importance. One can constrain the redshift and repeat analysis to find the optimal bands for stellar mass measurements in the desired redshift range given the availability of $i-$band data.

One should note that these conclusions depend on the selection criteria of the sample. This method provides a powerful tool in designing future surveys and quantifying the importance of each waveband. An efficient observation can be conducted by prioritizing important wavebands identified by the information gain-based method. 

\begin{figure}
    \centering
    \includegraphics[width=0.5\textwidth]{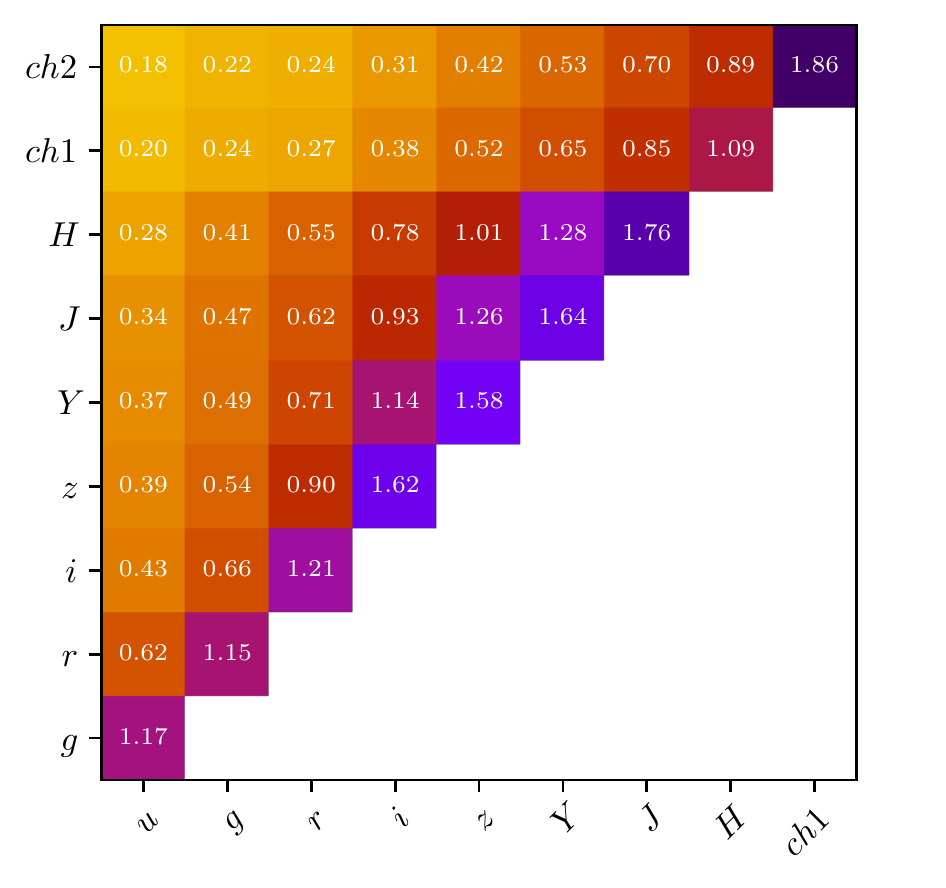}
    \caption{Visual representation of the mutual information between different wavebands for a sample of $i<25$ AB mag galaxies. The map is colored based on the value of mutual information, with purple representing the most correlated bands and yellow representing the least correlated bands (mostly independent). For instance, the mutual information of $ch1$ and $ch2$ quantifies the bits of information about IRAC/$ch1$ flux obtained by observing IRAC/$ch2$ flux. It is similar to the correlation coefficient, but it is able to capture non-linear relationships.}
    \label{fig:mi_bands}
\end{figure}

Moreover, different waveband fluxes can be inter-correlated for a specific sample of galaxies. For instance, the top left panel in Figure \ref{Fig:cmi_mass} shows that IRAC/$ch1$ and $ch2$ provide a comparable amount of information for stellar mass measurements, which suggests that these bands are inter-correlated for our sample with $i<25$ AB mag. Figure \ref{fig:mi_bands} visualizes the mutual information between different bands. A greater value of mutual information indicates that wavebands are more correlated. Inter-correlation between wavebands allows us to predict/simulate fluxes of galaxies in missing bands. In the following, we investigate the possibility of predicting/simulating near-IR UVISTA/$YJH$ fluxes based on H20-like data for a sample of galaxies with $i<25$ AB mag.

\vspace{1cm}
\section{Data Visualization} 
\label{sec:Visualize}
Fluxes of galaxies in $N$ wavebands are used to measure the photometric redshifts and physical parameters of galaxies. For example, the H20-like data with $N=7$ bands occupy a 7-dimensional space, where the position of each galaxy is determined by its fluxes in 7 bands. Therefore, galaxies with similar positions in $N$-dimensional space are expected to have similar redshifts and physical parameters if $N$ is large enough to fully sample the observed SED of galaxies. Similarly, it is expected that they will have similar fluxes in $(N+1)^{th}$ waveband. However, showing galaxy fluxes in a high-dimensional space (e.g., 7-dimensional space) is impossible and thus, we use dimensionality reduction techniques to present them in 2D space such that the information of higher dimension is maximally preserved. In this work, we use Uniform Manifold Approximation and Projection \cite[UMAP;][]{McInnes18} technique to visualize our sample in a 2-dimensional space. UMAP is a non-linear dimensionality reduction technique that estimates the topology of the high-dimensional data and uses this information to construct a low-dimensional representation of data that preserves structure information on local scales. It also outperforms other dimensional reduction algorithms such as t-SNE \citep[t-Distributed Stochastic Neighbor Embedding;][]{vanDerMaaten2008} used in the literature \citep{Steinhardt20} since it preserves structures on global scales as well. In a simple sense, UMAP constructs a high-dimensional weighted graph by extending a radius around each data point and connecting points when their radii overlap. This radius varies locally based on the distance to the $n^{th}$ nearest neighbor of each point. The number of the nearest neighbor (n) is the hyper-parameter in UMAP that should be fixed to construct the high-dimensional graph. Small (large) values for n will preserve more local (global) structures. Once the high-dimensional weighted graph is constructed, UMAP optimizes the layout of a low-dimensional map to be as similar as possible to the high-dimensional graph.             

\begin{figure}
    \centering
    \includegraphics[width=\linewidth]{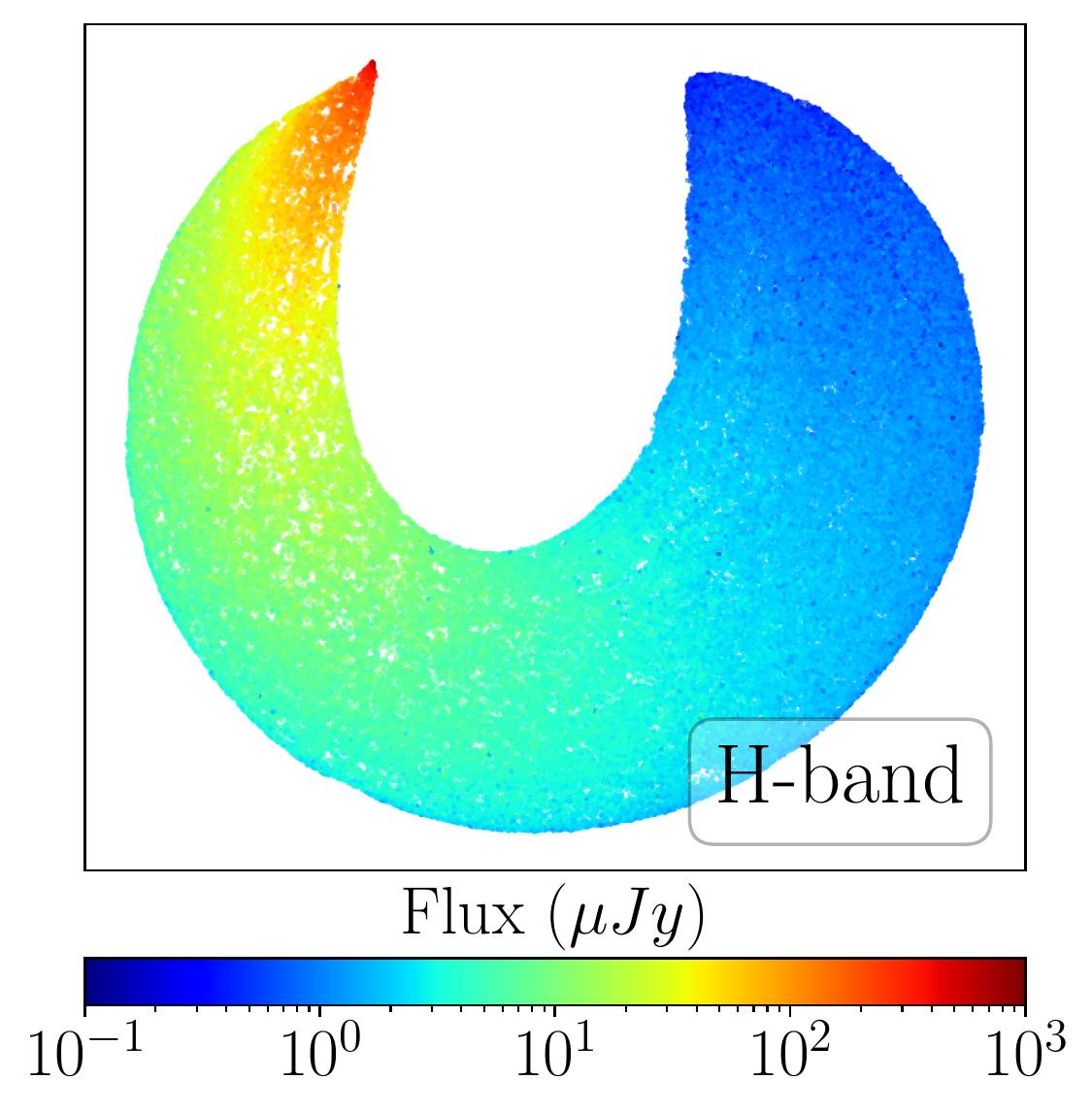}
    \caption{2-D visualization of the sample with H20-like bands using the UMAP technique. The mapped data are color-coded by the $H-$band fluxes. The smooth gradient of $H-$band fluxes in the 2-D representation reassures us that galaxies with similar fluxes in H20-like bands have similar $H-$band fluxes as well.}
    \label{fig:H_umap}
\end{figure}

We use the UMAP Python library\footnote{https://github.com/lmcinnes/umap} to map 7-dimensional flux space of H20-like data to 2 dimensions considering 50 nearest neighbors to provide a balance between preserving local and global structures. We do not map magnitudes or colors since non-detected values cannot be handled properly when using them. Multi-waveband fluxes contain all the information regarding colors, but using colors misses information regarding fluxes or magnitudes. Therefore, mapping fluxes of galaxies from that space to 2-dimension is a better way than using colors. Since fluxes in different bands have fairly similar distributions, no normalization is needed before applying UMAP. In the case of significantly distinct distributions, normalization is needed to avoid the dominance of a waveband with a larger dynamic range. Figure \ref{fig:H_umap} shows a 2-D visualization of the sample with H20-like bands using the UMAP algorithm. As an example, the mapped data are color-coded by the $H-$band fluxes (not present in H20 photometry) in $\mu{\rm Jy}$. The smooth transition of the $H-$band fluxes in the 2D representation in Figure \ref{fig:H_umap} reassures us that galaxies with similar fluxes in H20-like bands also have similar $H-$band fluxes. We note that the H20-like data set does not include $H-$band data.

Visualized data in Figure \ref{fig:H_umap} show qualitatively that the $H-$band fluxes are predictable to some extent using H20-like data. To perform a quantitative assessment on how accurately one can predict fluxes in UVISTA $YJH$ bands given the H20-like observations, we train a Random Forest \cite[][]{Breiman01} model with half of our sample and evaluate the model's performance with the other half. A Random Forest consists of an ensemble of regression trees. The algorithm picks a subsample of the dataset, builds a regression tree based on the subsample and repeats this procedure numerous times. The final value is the average of all the values predicted by all the trees in the forest. Having numerous decision trees based on subsampled data makes this algorithm unbiased and unaffected by overfitting. Another advantage of this method is that the inputs do not need to be scaled before feeding into the model. In the following section, we train a Random Forest model and evaluate its accuracy.    

\section{Flux predictions } 
\label{sec:band_prediction}

\begin{figure*}[]
    \centering
    \subfloat{{\includegraphics[width=8.55cm]{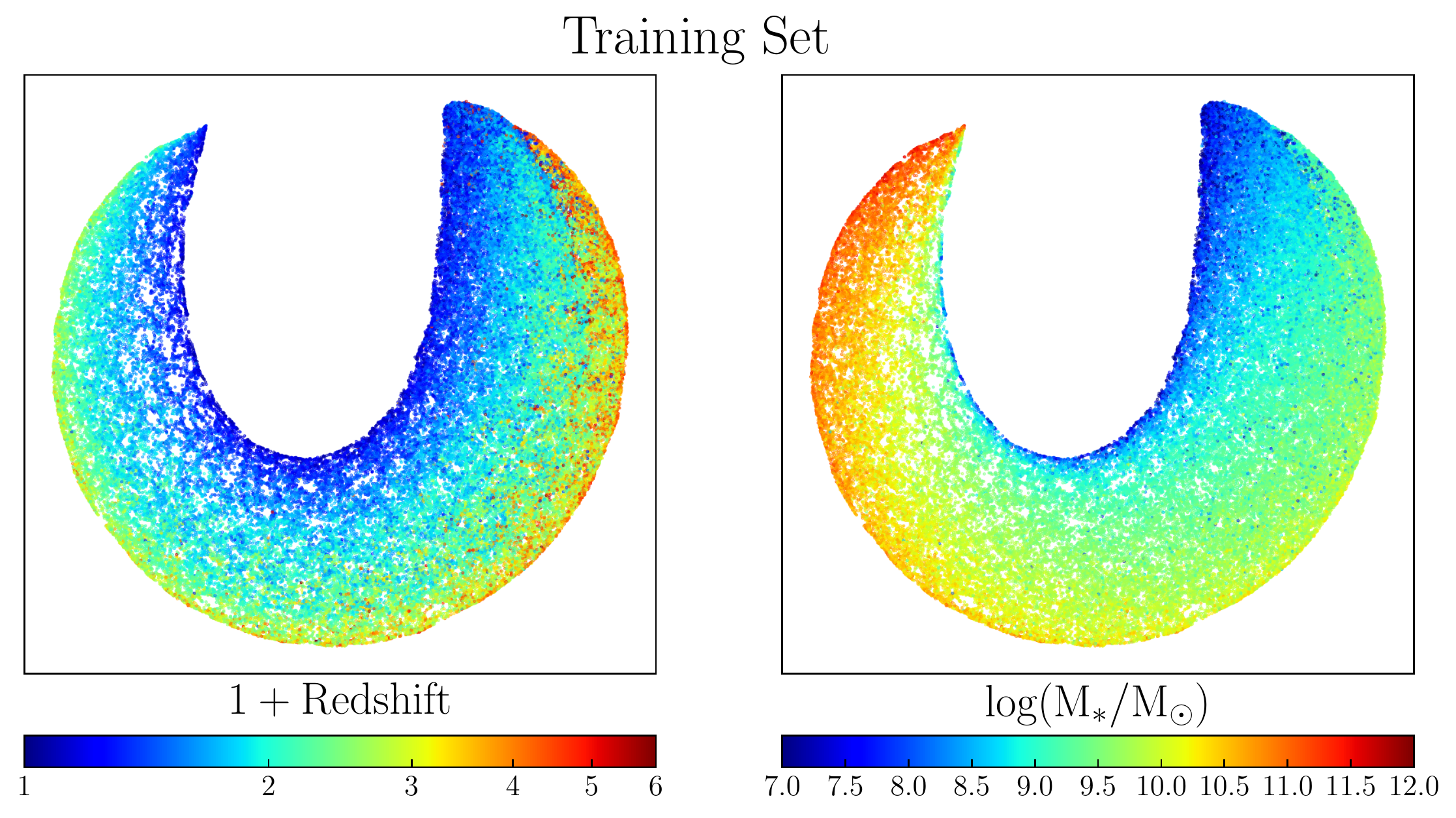} }}%
    \qquad
    \subfloat{{\includegraphics[width=8.55cm]{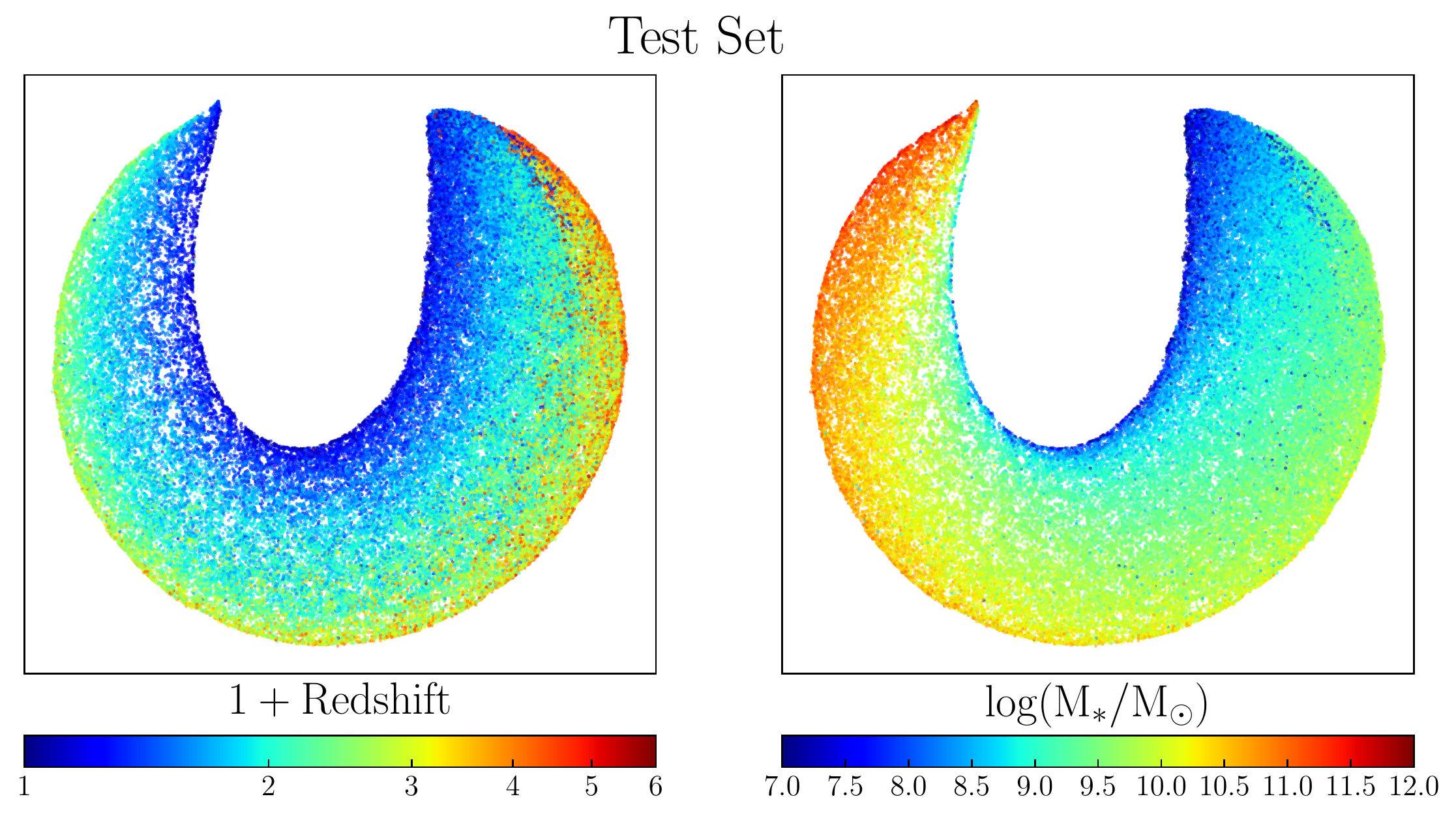} }}%
    \caption{\new{Similar to Figure \ref{fig:H_umap}, but for training (two left panels) and test (two right panels) samples. Maps are color-coded with photometric redshifts and stellar masses. We find that the training and test samples share the same properties, so the randomly selected training sample is representative of the galaxies in the COSMOS field.}  }%
    \label{fig:umap_train_test}%
\end{figure*}

\begin{figure}
    \centering
    \subfloat{{\includegraphics[width=0.9\columnwidth, clip=True, trim=0cm 0.25cm 0cm 0.25cm]{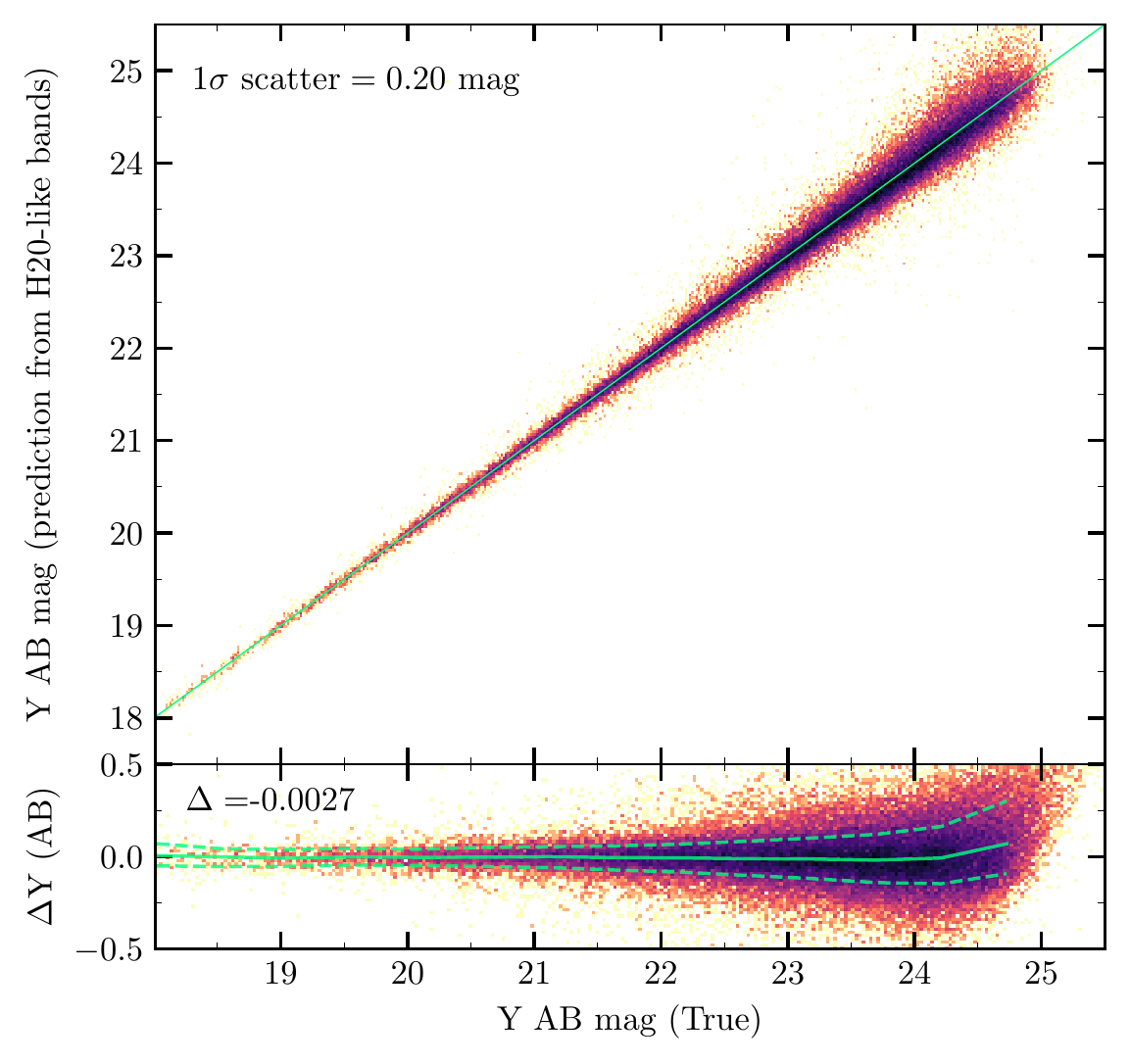} }}%
    
    \subfloat{{\includegraphics[width=0.9\columnwidth, clip=True, trim=0cm 0.25cm 0cm 0.25cm]{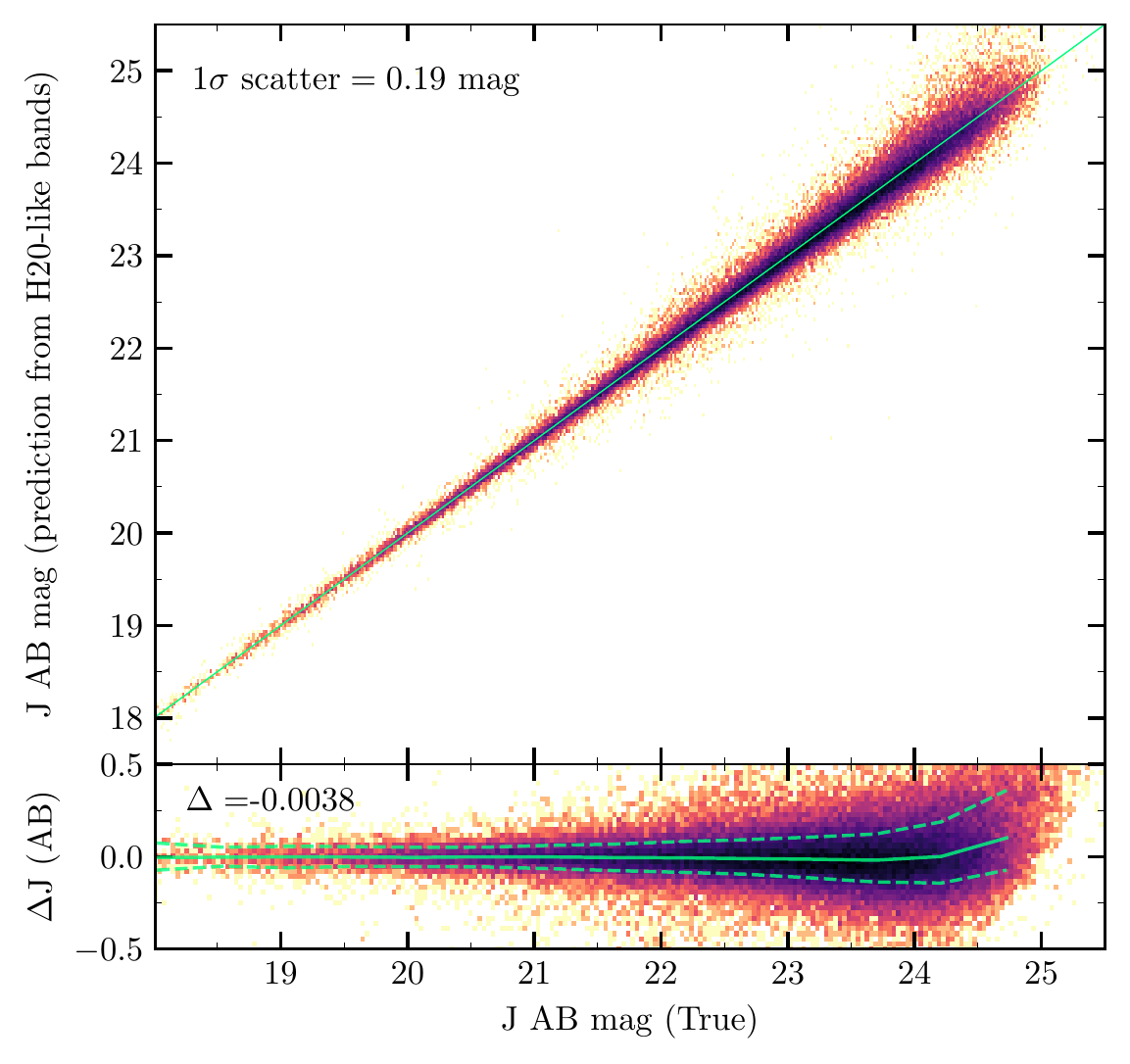} }}%
    
    \subfloat{{\includegraphics[width=0.9\columnwidth, clip=True, trim=0cm 0.25cm 0cm 0.25cm]{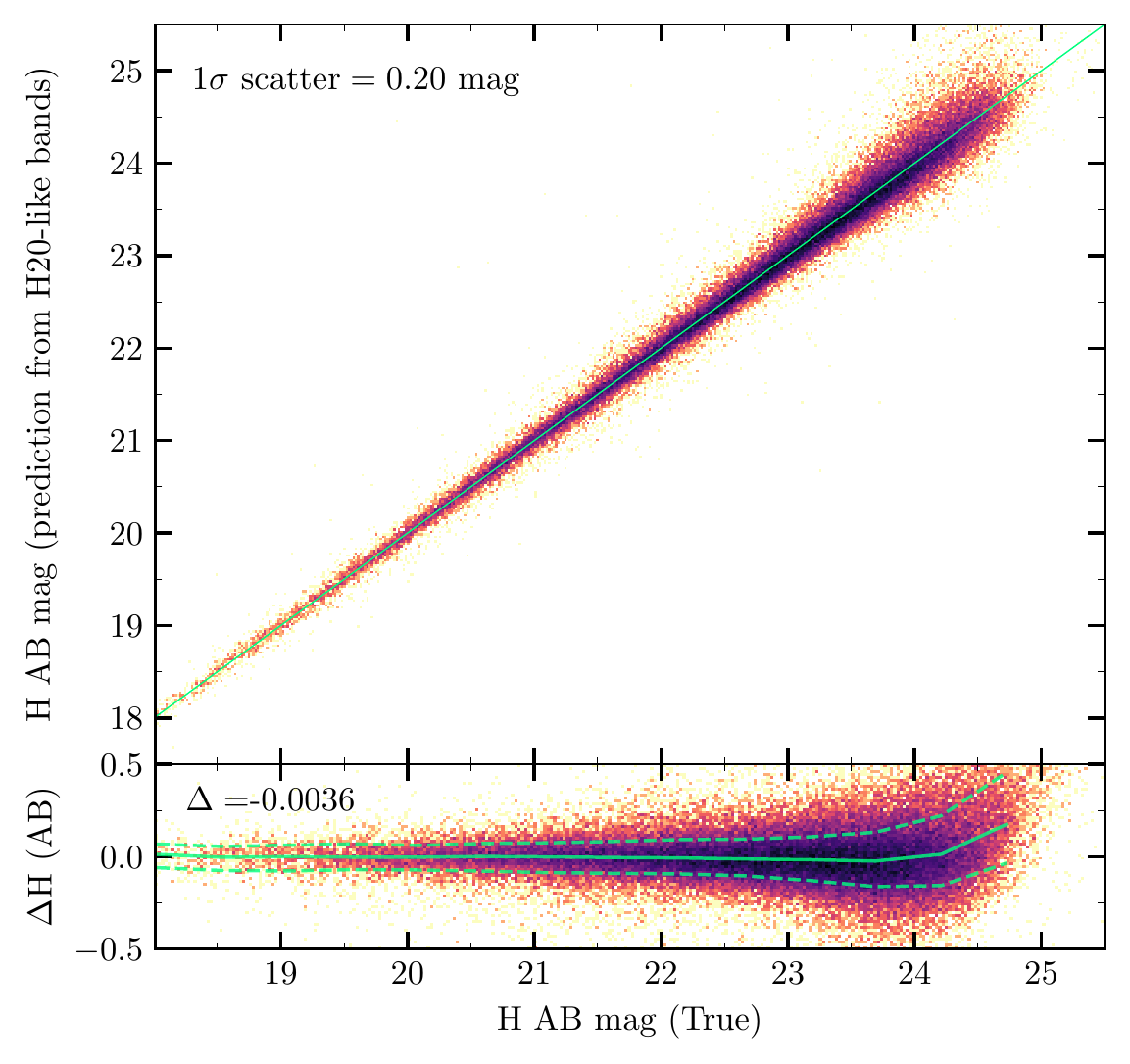} }}%
    \caption{The performance of the Random Forest model on the 82,904 test galaxies not used for the training of the model. The model is trained based on H20-like bands ($u,g,r,i,z,ch1,ch2$) and predicts UVISTA $YJH$ bands. Bottom panels show the scatter of $\rm Mag_{Predicted}-Mag_{True}$ as a function of true magnitudes and $\Delta$ is the median offset in these scatter plots.}%
    \label{fig:Euclid_RF}%
\end{figure}

We split the sample (described in Section \ref{sec:Data}) \new{randomly} into a training and a test sample. \new{To evaluate if the training sample is representative, we construct a 2-D projection of H20-like fluxes similar to Figure \ref{fig:H_umap} for both training and test samples. Figure \ref{fig:umap_train_test} shows the 2-D visualizations color-coded by the properties of galaxies (photometric redshift and stellar mass). We find that the training and test samples share the same properties, so the training sample is representative of the galaxies in the COSMOS field.} With 82,903 galaxies as a training sample, we build a Random Forest model with 100 regression trees to predict UVISTA $YJH$ bands from the H20-like band fluxes. We use Python implementation of the algorithm \cite[Scikit-learn;][]{scikit-learn} \footnote{https://scikit-learn.org/stable} with its default parameters to build the model. The true (observed) fluxes in the $YJH$ bands are available in the COSMOS2020 catalog. Using the trained Random Forest model, we then predict the expected fluxes for galaxies not included in the training set, with the results compared in Figure \ref{fig:Euclid_RF}. For each band, we compare the predicted magnitudes ($\rm Mag_{Predicted}$) with the true observed magnitudes ($\rm Mag_{True}$). We find that the Random Forest model predicts unbiased $YJH$ fluxes with high accuracy. The bottom panel in each figure shows the scatter of the $\rm Mag_{Predicted}-Mag_{True}$ as a function of true magnitudes. With a median magnitude discrepancy ($\Delta$) of $\sim 0.01$, we find that the offset is comparable with discrepancies that arise from different methods of photometric data reduction. \cite{Weaver21} found that the median tension between the magnitudes derived from aperture photometry and profile-fitting extraction is $\Delta\sim 0.002$ in $YJ$ bands and $\Delta\sim 0.02$ in $H-$band for sources brighter than the 3$\sigma$ depth of each band. Thus, such small offsets in the Random Forest regressor are within the intrinsic uncertainties of the data reduction techniques. Green solid and dashed lines in the sub-panels of Figure \ref{fig:Euclid_RF} show the median of $\Delta$ and 1$\sigma$ (68\%) scatter, respectively. The scatter in the prediction is $<0.17$ mag for galaxies brighter than 24 AB mag. This shows that $YJH$ near-IR observations of UVISTA can be simulated with acceptable accuracy from the available observations of H20 for a sample of galaxies with $i<25$ AB mag. \new{Our results remain consistent when we rebuild a new Random Forest with different randomly selected training samples.} While our focus in this paper is on the UVISTA/$YJH$ and H20 bands, the method we present is general and directly applicable to other surveys.

\section{Photometric redshift and stellar mass} 
\label{sec:Photz-M}

\begin{figure*}[]
    \centering
    \subfloat{{\includegraphics[width=8.5cm]{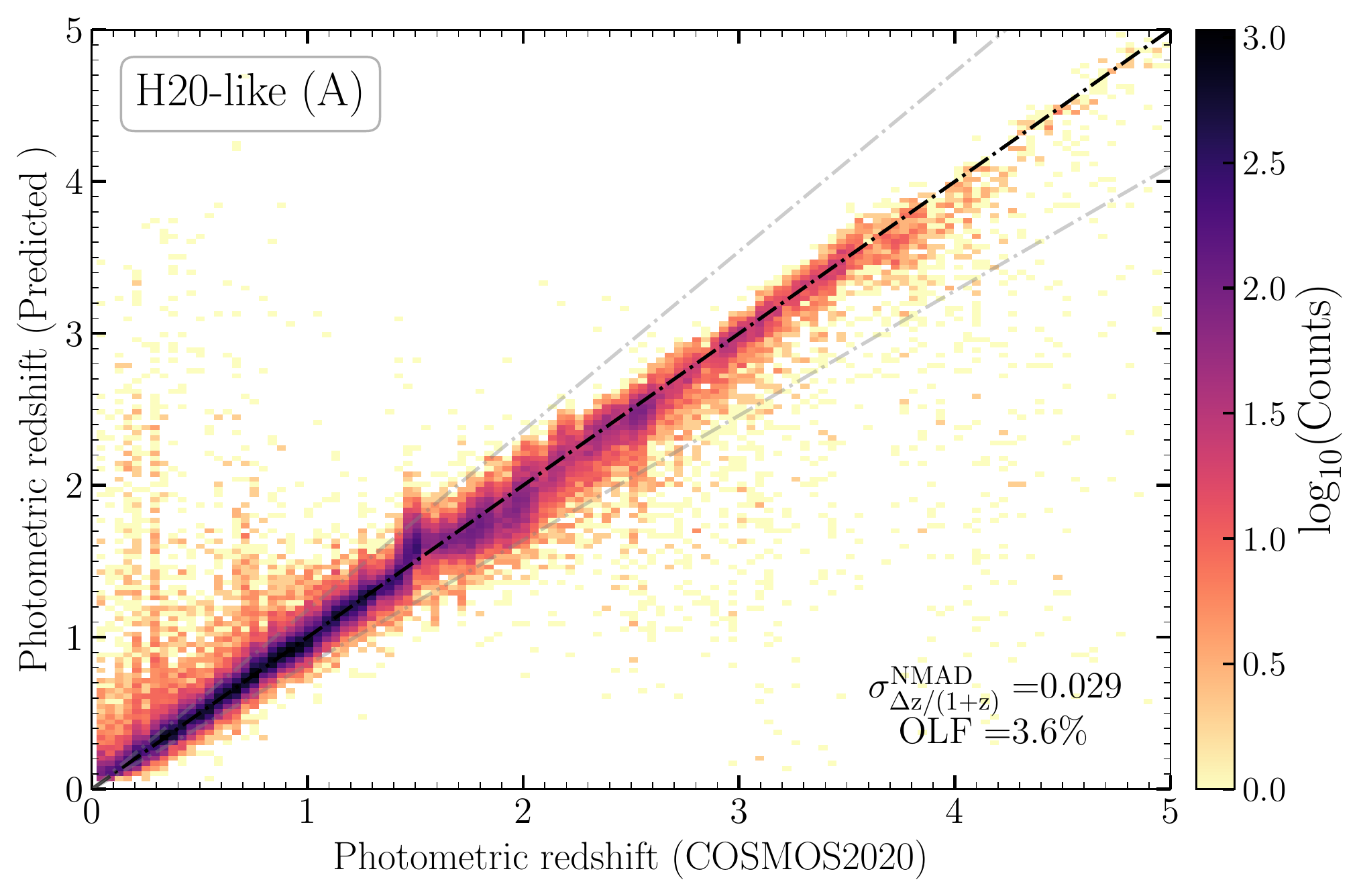} }}%
    \qquad
    \subfloat{{\includegraphics[width=8.5cm]{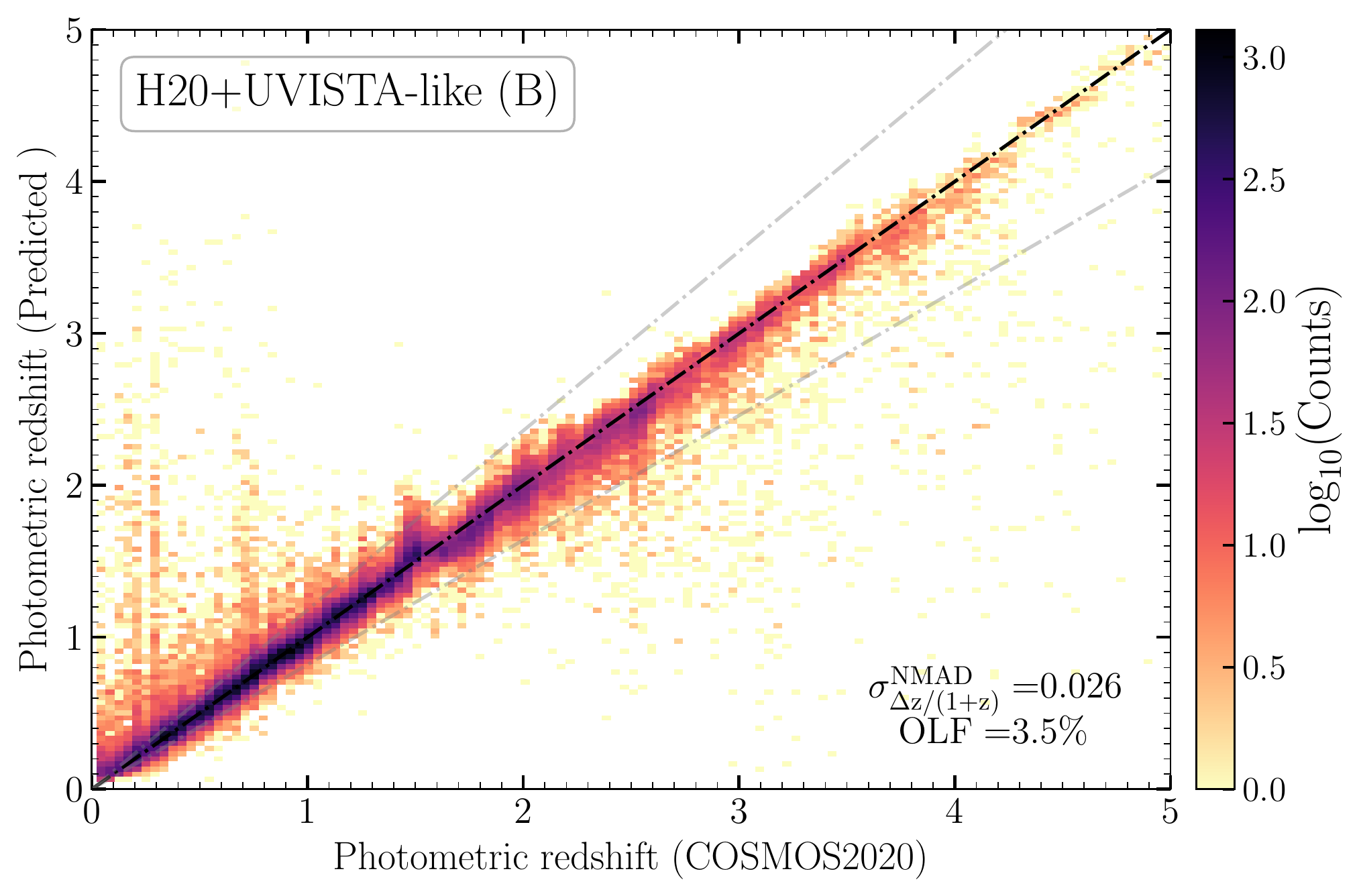} }}%
    \qquad
    \subfloat{{\includegraphics[width=8.5cm]{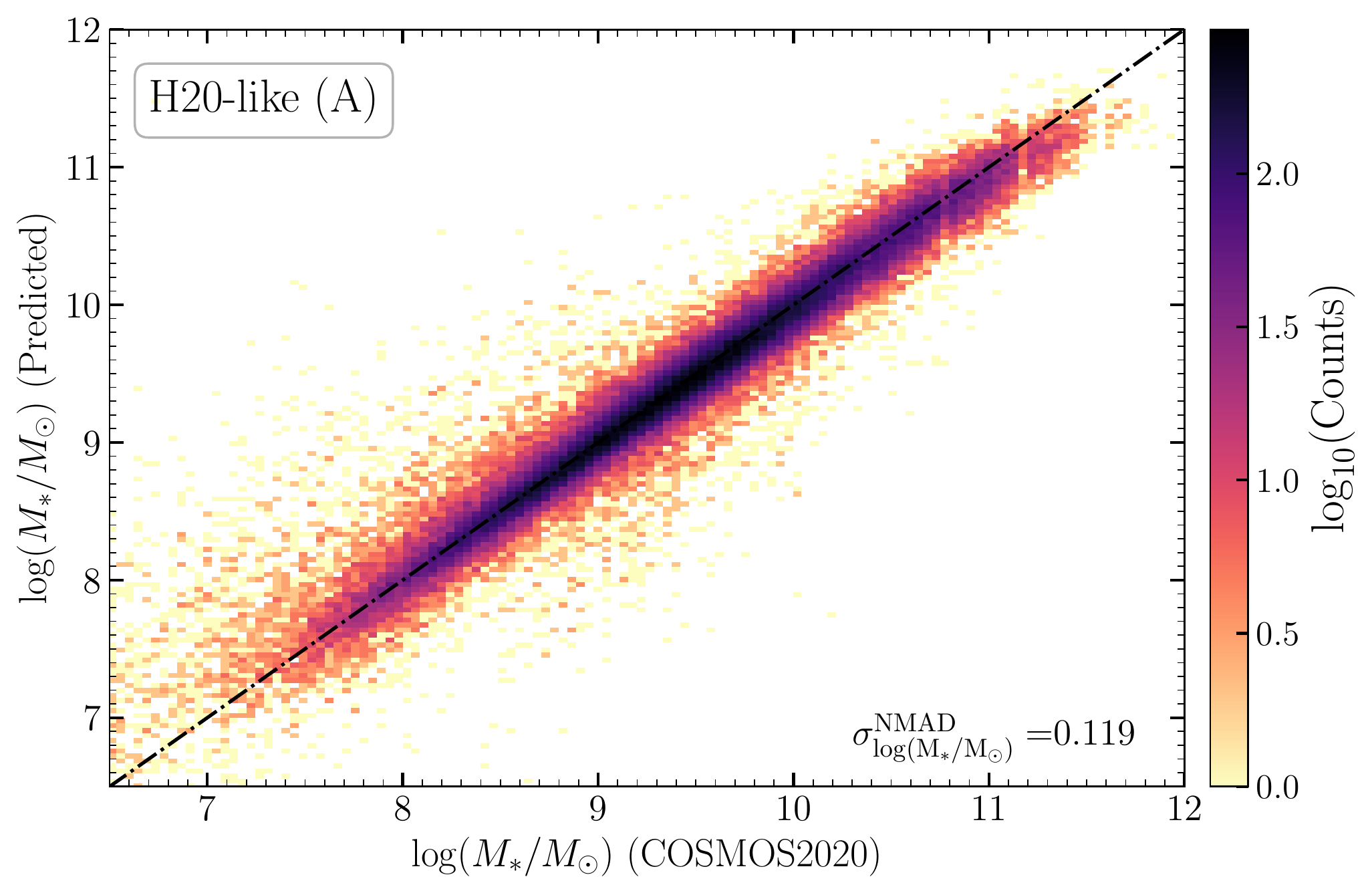} }}%
    \qquad
    \subfloat{{\includegraphics[width=8.5cm]{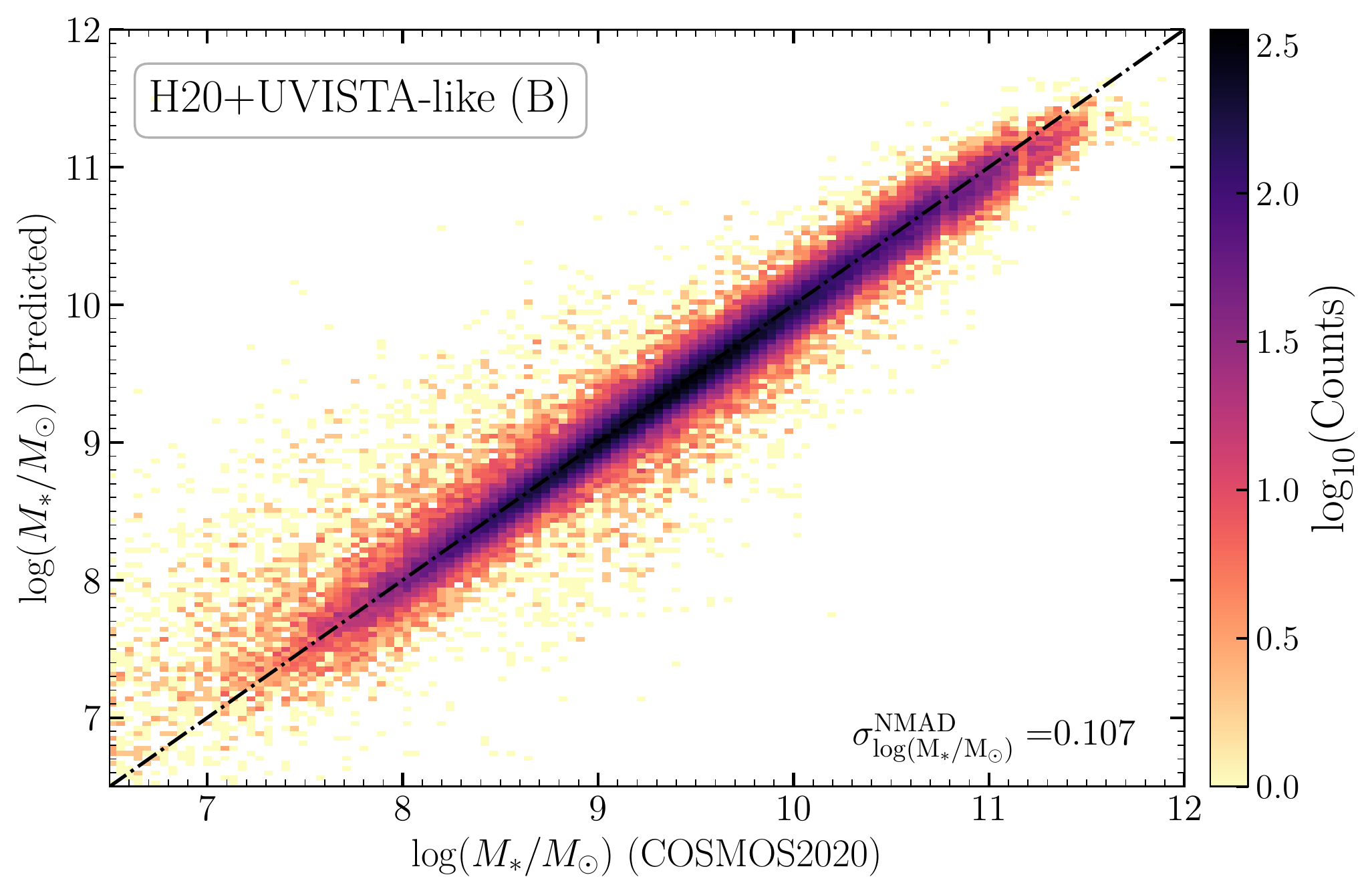} }}%
    \caption{Performance of the Random Forest model to predict photometric redshifts and stellar masses when the model is trained by H20-like bands (left panels) and H20+UVISTA-like bands (right panels). Both trained models recover photometric redshifts and stellar masses with high accuracy. The similar performance of the model with and without $YJH$ bands originates from the fact that the H20-like bands capture most of the information available in $YJH$ bands as shown in Figure \ref{fig:Euclid_RF}. The black dashed lines show one-to-one relation, and the gray dashed lines correspond to the predicted redshifts at $\pm0.15(1+z)$ (outlier definition boundaries).}%
    \label{fig:z_RF}%
\end{figure*}

\begin{figure*}
    \centering
    \includegraphics[width=\textwidth]{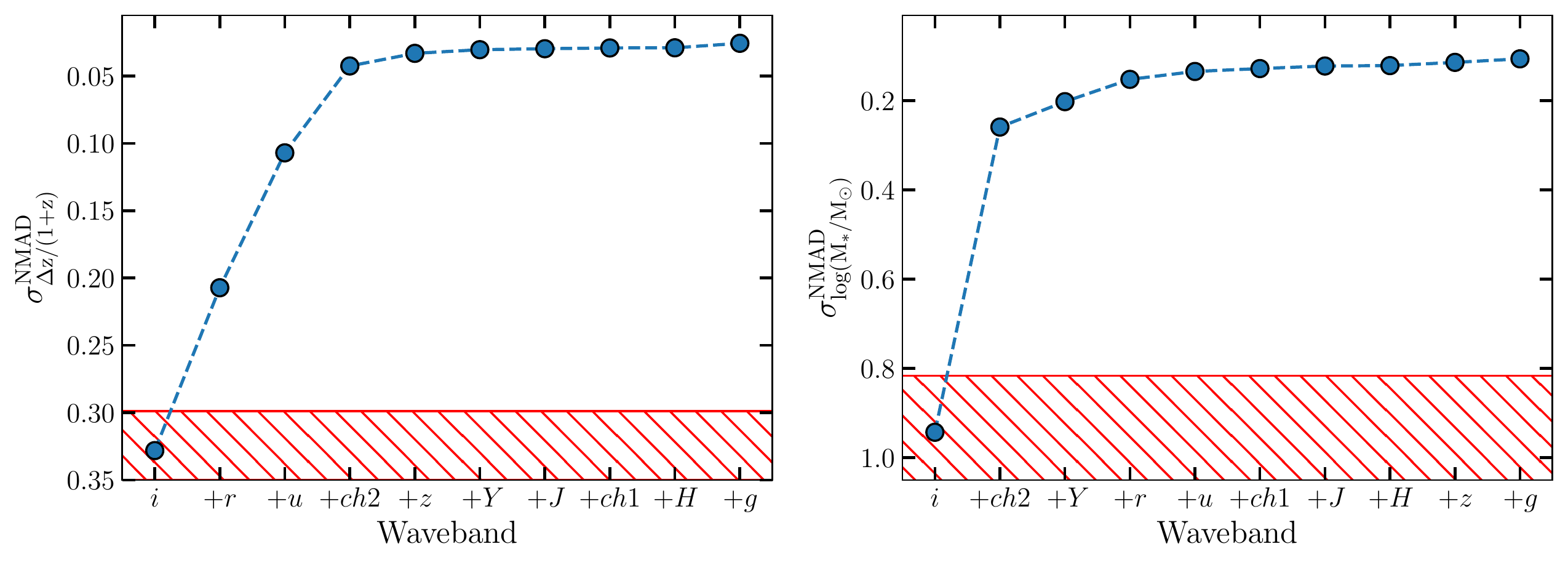}
    \caption{The normalized median absolute deviation of $\Delta z/(1+z)$ (left) and $\log(M_*/M_\odot)$ (right) as a function of bands used to measure the parameter. As the sample is selected based on the $i-$band magnitude of galaxies, we start with training a Random Forest model based on only $i-$band data and then we add other bands following the same order of importance we find in Figure \ref{Fig:cmi} and \ref{Fig:cmi_mass}. \new{Red horizontal lines show the scatter of the data relative to their mean value.  }    }
    \label{fig:RF_comp}
\end{figure*}

In the previous section, we showed that given the observations of the H20 survey, near-IR observations of UVISTA can be constrained to some extent. In other words, observations of the COSMOS field provide valuable information regarding the distribution of galaxies in the flux space, even if we do not observe galaxies as extensively as it is done in the COSMOS field in terms of spectral coverage. When we use template fitting code with synthetic templates, we usually do not take into account this constraint. There are two approaches to incorporate this information in the photometric redshifts or physical parameters measurements. First, add a prior to fluxes in the bands that are not observed in the survey. For instance, when we perform SED fitting using H20-like bands, we can add priors to the $YJH$ bands based on a Random Forest model, which is trained over the population of galaxies from the COSMOS observations. Second, train a model based on SED-fitting results calculated with a large number of bands. In this case, when we feed our model with H20-like data, it will decide the best value of a parameter based on both the existence of similar observations in the COSMOS field (information from galaxy populations) and the SED-fitting solution for that galaxy. 

In this section, we employ the latter approach to train a model to predict the photometric redshifts and the stellar masses of galaxies based on H20-like and H20+UVISTA-like bands. We train a Random Forest model based on a training sample of observed galaxies. The inputs of the model are H20-like fluxes and the output is either photometric redshift or stellar mass computed from SED fitting over 29 bands available in the COSMOS2020 catalog. We also train another similar model where the inputs are H20+UVISTA-like bands. Figure \ref{fig:z_RF} shows the performance of trained models on the test sample with 82,904 galaxies. We find that both models recover photometric redshifts and stellar masses with comparable accuracy with being slightly accurate using H20+UVISTA-like inputs. Normalized median absolute deviation ($\sigma_{\rm NMAD}$) of $\Delta z/(1+z)$ is $\sim 0.03$ for both models with $\sim 4\%$ outlier fraction. Outlier galaxies are defined as galaxies with $\Delta z/(1+z)>0.15$. The median absolute deviation of $\log (M_*/M_\odot)$ is $\sim 0.1$ dex for both models. We explain this similar performance using the results of Section \ref{information} and \ref{sec:band_prediction}. The Random Forest model with H20-like bands comprises most of the information regarding UVISTA bands as we trained the model with the population of observed COSMOS galaxies. Therefore, it should recover photometric redshifts and stellar masses as accurately as the model which includes near-IR ($YJH$) observations. 

We repeat a similar analysis starting with only $i-$band data and adding other important bands in the same order as we identified in Section \ref{information}. Figure \ref{fig:RF_comp} shows the the normalized median absolute deviation of $\Delta z/(1+z)$ and $\log(M_*/M_\odot)$ as a function of bands used to measure the parameter. We find that $i-,r-,u-,ch2-,z-$band are the minimal number of bands to reach an acceptable accuracy of $\sigma^{\rm NMAD}_{\Delta z/(1+z)}=0.03$ to measure photometric redshifts of $i<25$ AB mag. For the same sample, $i-,ch2-,Y-,r-,u-$band are the optimal bands for stellar mass measurements reaching an accuracy of $\sigma^{\rm NMAD}_{\log(M_*/M_\odot)}=0.15$ dex.

\subsection{Synthetic templates}

In the following, we use UMAP to visualize photometry of synthetic SED models commonly used in template-fitting procedures. We build a set of theoretical templates using 2016 version of a library of \cite{Bruzual03}, considering \cite{Chabrier03} initial mass function. Star formation histories are modeled with an exponentially declining function (${\rm SFR} \propto e^{-t/\tau}$), where $\tau$ is the star formation timescale. Dust attenuation is applied using the \cite{Calzetti} law and solar stellar metallicity is assumed for all templates. We build $\sim750,000$ theoretical templates assuming $\tau\in(0.1,10)\ \rm Gyr$, $t\in(0.1,13.7)\ \rm Gyr$, $ A_V\in(0,2)\ \rm mag$ and $z\in(0,5.5)$. $t$ and $A_V$ are the stellar age and the extinction in the visual band, respectively. We then calculate the synthetic photometry in both H20-like and H20+UVISTA-like bands by applying the corresponding filter response function. 

\begin{figure}
    \centering
    \includegraphics[width=\linewidth]{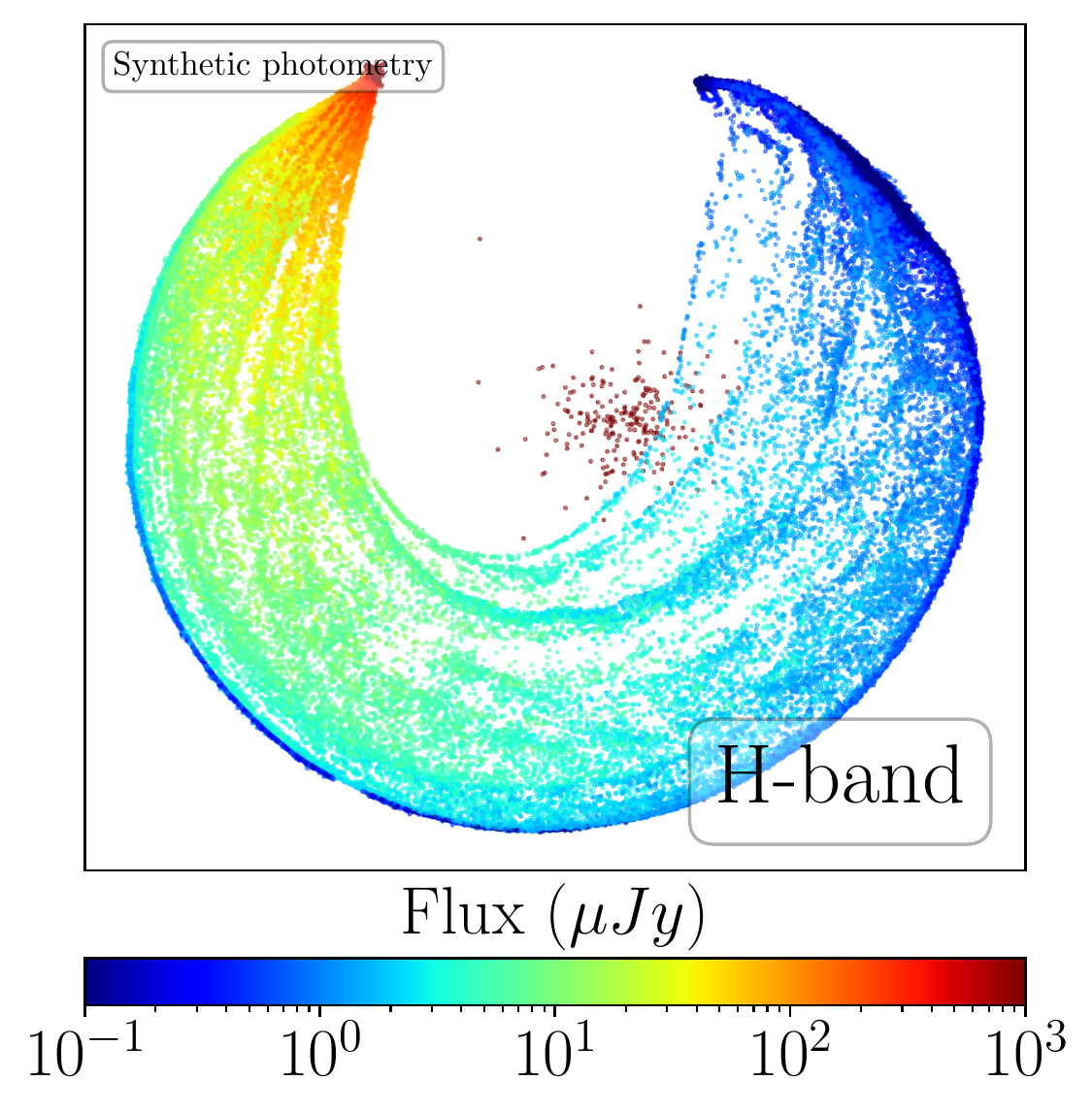}
    \caption{Similar to Figure \ref{fig:H_umap}, but for synthetic photometric data. The high-dimensional synthetic H20-like data are transformed to the space learned in Figure \ref{fig:H_umap}. The map is color-coded by the synthetic $H-$band fluxes. Existing dissimilarities between this figure and Figure \ref{fig:H_umap} show that synthetic models lack the observed information.}
    \label{fig:H_umap_synthetic}
\end{figure}

As we learned the topology of fluxes in the H20-like bands for real observed galaxies in COSMOS2020 catalog (Figure \ref{fig:H_umap}), we can transform H20-like band fluxes of synthetic photometry into the learned space. Figure \ref{fig:H_umap_synthetic} shows the 2-D visualization of the theoretical templates with H20-like bands in that learned space. As an example, data points in the reduced dimension are color-coded by their synthetic $H-$band fluxes in $\mu{\rm Jy}$. Comparing theoretical templates with the observed data shown in Figure \ref{fig:H_umap} reveals that model galaxies encounter degeneracies. In this specific example, we show that templates with similar H20-like fluxes have more diverse $H-$band fluxes than real observations, which can produce degenerate results when template fitting is performed based on H20-like bands. Adding information of the COSMOS2020 observations as a prior imposes a strong correlation between the observed and missing bands and makes the theoretical templates less degenerate as shown in Figure \ref{fig:H_umap}. For example, the dark blue arc in the left side of Figure \ref{fig:H_umap_synthetic} mismatches with the observational counterpart. In other words, synthetic templates predict $H-$band flux of $\sim 0.1$ $\mu{\rm Jy}$ for galaxies in that vicinity (i.e., the dark blue arc), but real observations show that they have, in fact, $H-$band flux of $\sim 10$ $\mu{\rm Jy}$. This shows that extra information that exists in the previous observations can add valuable information to template fitting analysis. 

If one adds a predicted band in the template-fitting procedure, the errors should be assigned based on the $1\sigma$ scatter of the predicted flux (dashed green lines in Figure \ref{fig:Euclid_RF}). It is particularly important to properly take into account the systematic scatter of the predicted bands in template-fitting and ensure that the predicted bands are not over-weighted in best-template selection. \new{In the following section, we perform a simple template-fitting to evaluate values added by predicted fluxes.} However, it is worth highlighting that the better approach would be using a machine learning model which is trained based on template-fitting results of a galaxy population with well-constrained SEDs such as COSMOS2020 (Figure \ref{fig:z_RF}).

\subsection{Template-fitting}
\label{sec:Template-fitting}

\begin{figure*}[]
    \centering
    \subfloat{{\includegraphics[width=\linewidth]{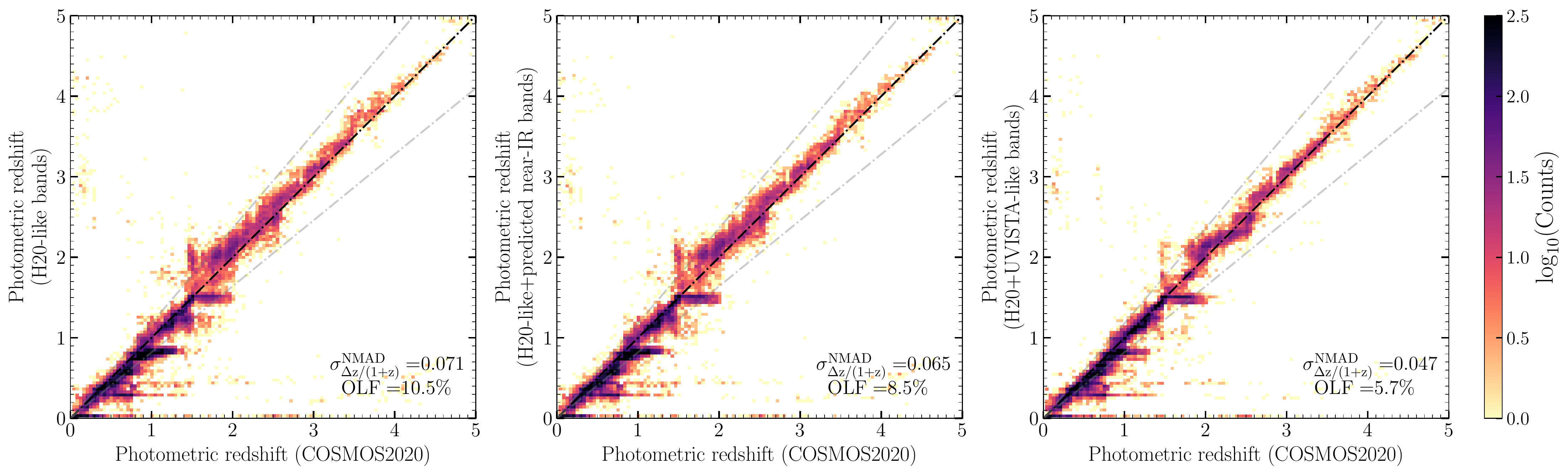} }}%
    \qquad
    \subfloat{{\includegraphics[width=\linewidth]{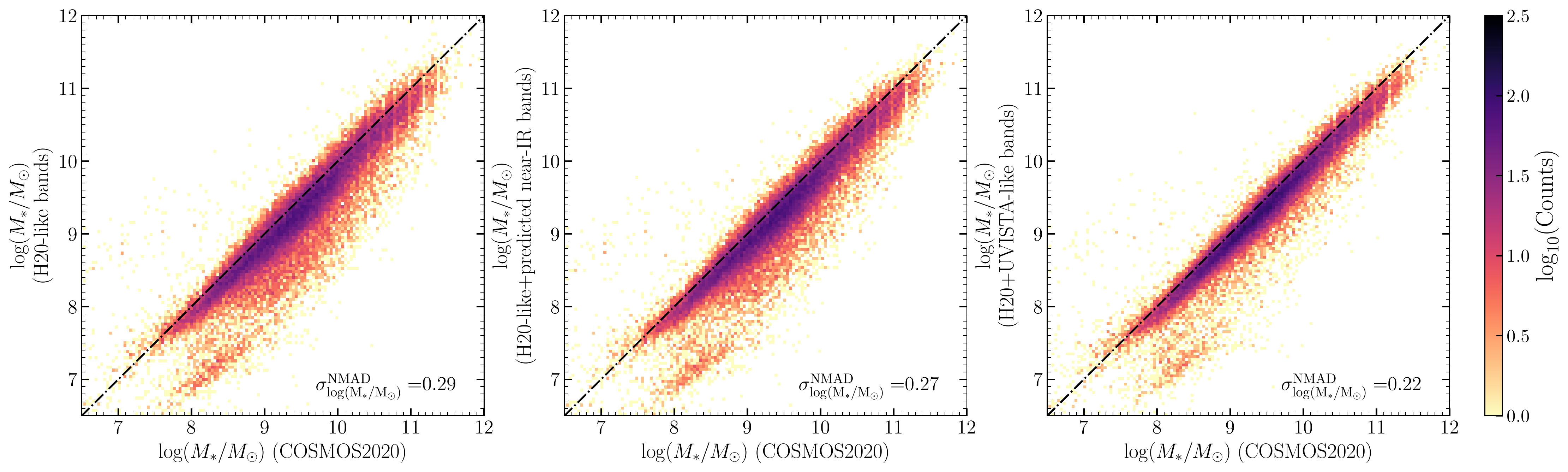} }}%
    \caption{\new{Template-fitting results are compared against photometric redshifts and stellar masses of COSMOS2020 catalog (derived from 29 bands) for three cases: using 1) observed ugrizch1ch2 bands (left panels), 2) observed ugrizch1ch2+predicted JHK bands (middle panels) and 3) observed ugrizYJHch1ch2 bands (right panels).    }}%
    \label{fig:z_SED}%
\end{figure*}

\new{We perform template fitting for three cases using 1) H20-like bands, 2) H20-like+predicted YHJ bands, and 3) H20+UVISTA-like bands. For this purpose, we split the test sample used in Section \ref{sec:band_prediction} into half to have a validation set as well as a new test sample. The validation sample is used to measure $1\sigma$ scatter of the predicted flux (similar to dashed green lines in Figure \ref{fig:Euclid_RF}). We assign errors to the predicted fluxes of the new test sample based on $1\sigma$ scatter of the validation sample at a given magnitude. We use a template-fitting code \lephare{} with the same configuration as \cite{Ilbert15}. This configuration differs from the templates used for COSMOS2020 redshift measurements. In the COSMOS2020 catalog, the photometric redshifts are measured based on templates employed by \cite{Ilbert13}, followed by stellar masses measured in the same manner as \cite{Ilbert15} at fixed photometric redshifts, but here we fit both photometric redshifts and stellar masses simultaneously. Figure \ref{fig:z_SED} presents the results of the template-fitting for these three cases. We find that the lack of observed near-IR fluxes in template-fitting increases the $\sigma_{\rm NMAD}$ and outlier fraction by 50\% and 80\%, respectively. We also find that adding predicted fluxes improves the $\sigma_{\rm NMAD}$ and outlier fraction by 10\% and 25\%, respectively. Predicted fluxes also improve the scatter of the stellar mass measurements by 7\%.} 

\new{Improvement in template-fitting results by adding predicted fluxes suggests that observationally driven priors on near-IR fluxes can help reduce both scatter and outlier fraction of SED-derived properties. Moreover, we find that adding observed near-IR data significantly ($\sim50\%$) improves the template-fitting results, but this is not the case for the Random Forest model shown in Figure \ref{fig:RF_comp} ($\sim10\%$ improvement). This suggests that machine learning models are able to fully incorporate the information gathered from extensive surveys and avoid the degeneracies in template-fitting parameters that are inevitable when a few bands are present.}
\section{Discussion and Summary}
\label{sec:Discussion_Summary}

In this paper, we present an information gain-based method to quantify the importance of wavebands and find the optimal set of bands needed to be observed to constrain photometric redshifts and physical properties of galaxies. To demonstrate the application of this method we build a subsample of galaxies from COSMOS2020 catalog with similar waveband coverage ($ugrizYJH$ and IRAC/$ch1,ch2$) that will be available in \Euclid{} deep fields. For a sample of galaxies with $i<25$ AB mag, we find that given the availability of $i$-band fluxes, $r, u$, IRAC/$ch2$ and $z$ bands provide most of the information for measuring the photometric redshifts with importance decreasing from $r-$band to $z-$band. We also find that for the same sample, IRAC/$ch2$, $Y$, $r$ and $u$ bands are the most relevant bands in stellar mass measurements with decreasing order of importance. We note that these results should be remeasured for any new sample with different selection criteria. Moreover, we present the relative importance of wavebands for stellar mass measurements in the bins of redshifts since their importance depends on the redshift. We also investigate the inter-correlation between the flux in different wavebands and use a machine learning technique to predict/simulate missing fluxes from a survey. To prove the concept, we apply the method trained on the COSMOS2020 data to predict UVISTA near-IR observations based on H20-like survey data, which include $ugriz$ and Spitzer/IRAC observations. We find that near-IR bands ($YJH$) can be predicted/simulated from ground-based ($ugriz$) and mid-IR Spitzer (IRAC/$ch1,ch2$) observations with an accuracy of $1\sigma$ mag scatter $\lesssim 0.2$ for galaxies brighter than $24$ AB mag in near-IR bands. We demonstrate that theoretical templates lack such valuable information already observed through numerous bands in the COSMOS field. We conclude that degeneracies in template-fitting can be alleviated if one trains a model based on template-fitting solutions for observed galaxies with extensive observations instead of using conventional SED fitting. We show that a model trained on H20-like bands has comparable accuracy to a model which is trained over H20+UVISTA-like bands, given that the model is trained over the observed galaxy population with a vast number of wavebands. 

\cite{Masters15} mapped high-dimensional color space of COSMOS galaxies in UVISTA bands using the self-organizing map (SOM) technique \citep{Kohonen1982} and proposed a spectroscopy survey to fully cover regions in reduced color space with no spectroscopic redshifts. This survey, C3R2, was awarded 44.5 nights on Keck telescope to map the color-redshift relation necessary for weak lensing cosmology \citep{Masters17,Masters19}. Later on, \cite{Hemmati19} used SOM to map the color space of theoretical models and used the reduced map as a fast template-fitting technique. \new{In the present work, we use a new technique, UMAP, to create a 2-dimensional representation of a high-dimensional flux distribution. This technique can also be utilized to map the color space of galaxies and study their physical properties (similar to Figure \ref{fig:umap_train_test}), providing an opportunity for further analyses that can be performed in the future.}

Acquiring data for galaxy surveys over wide areas and a range of wavelengths with a large number of wavebands is costly. A new method based on machine learning algorithms is presented in this paper to supplement the present and future surveys in their missing bands with information from previous extensive surveys (e.g. COSMOS). It can be used to optimize observations of future surveys, as well as to predict photometry of observatories that have ceased operation \citep{Dobbels20}. 

\section*{Acknowledgments} 
\new{We thank the anonymous referee for providing insightful comments and suggestions that improved the quality of this work.} NC and AC acknowledge support from NASA ADAP 80NSSC20K0437. ID has received funding from the European Union's Horizon 2020 research and innovation program under the Marie Skłodowska-Curie grant agreement No. 896225. 
\bibliography{Missing_bands}
\end{document}